\documentclass[pre,aps]{revtex4}

\usepackage{graphicx, subfigure}

\begin{document}

%%%%%%%%%%%%%%%%%%%%%%%%%%%%%%%%%%%%%%%%%%%%%%%%%%%%%%%%%%%%%%%%%%%%%
%%%%%%%%%%%   TITLE PAGE            %%%%%%%%%%%%%%%%%%%%%%%%%%%%%%%%%
%%%%%%%%%%%%%%%%%%%%%%%%%%%%%%%%%%%%%%%%%%%%%%%%%%%%%%%%%%%%%%%%%%%%%

\title{
       Data-driven derivation of the turbulent energy cascade
       generator
      }

\author{
Jochen Cleve$^{1,2}$, Thomas Dziekan$^{2,3}$, J\"urgen
Schmiegel$^{4}$, Ole E.\ Barndorff-Nielsen$^{4}$, Bruce R.\
Pearson$^{5}$, Katepalli R.\ Sreenivasan$^{1}$, and Martin
Greiner$^{6}$ }

\address{$^{1}$ICTP, Strada Costiera, 11, 34014 Trieste, Italy;
         email: cleve@ictp.trieste.it, krs@ictp.trieste.it}
\address{$^{2}$Institut f\"ur Theoretische Physik,
               Technische Universität,
               D-01062 Dresden, Germany}
\address{$^{3}$Department of Physics, Uppsala University,
               Box 530, S-75121 Uppsala, Sweden;
         email: Thomas.Dziekan@fysik.uu.se}
\address{$^{4}$Department of Mathematical Sciences, University of
               Aarhus, Ny Munkegade, DK-8000 Aarhus, Denmark;
         email: schmiegl@imf.au.dk, oebn@imf.au.dk}
\address{$^{5}$School of Mechanical, Materials,
               Manufacturing Engineering and Management,
               University of Nottingham,
               Nottingham NG7 2RD, United Kingdom;
         email: bruce.pearson@nottingham.ac.uk}
\address{$^{6}$Corporate Technology, Information{\&}Communications,
               Siemens AG, D-81730 M\"unchen, Germany;
         email: martin.greiner@siemens.com}

\date{18.12.2003}

\begin{abstract}
Within the framework of random multiplicative energy cascade
models of fully developed turbulence, expressions for two-point
correlators and cumulants are derived, taking into account a
proper conversion from an ultrametric to an Euclidean two-point
distance. The comparison with two-point statistics of the
surrogate energy dissipation, extracted from various wind tunnel
and atmospheric boundary layer records, allows an accurate
deduction of multiscaling exponents and cumulants. These 
exponents serve as the input for parametric estimates of the
probabilistic cascade generator.
\end{abstract}

\maketitle

%%%%%%%%%%%%%%%%%%%%%%%%%%%%%%%%%%%%%%%%%%%%%%%%%%%%%%%%%%%%%%%%%%%%%%%
%%%%%%%%%%%%%%%%%%%%%%%%%%%%%%%%%%%%%%%%%%%%%%%%%%%%%%%%%%%%%%%%%%%%%%%
\section{Introduction}
%%%%%%%%%%%%%%%%%%%%%%%%%%%%%%%%%%%%%%%%%%%%%%%%%%%%%%%%%%%%%%%%%%%%%%%
%%%%%%%%%%%%%%%%%%%%%%%%%%%%%%%%%%%%%%%%%%%%%%%%%%%%%%%%%%%%%%%%%%%%%%%

The inertial-range dynamics of fully developed turbulent flows
reveals an ubiquitous multiscale character, driven by large-scale
forcing on one end and controlled by fluid viscosity and
small-scale dissipation on the other. For the intermediate scale
range, the underlying Navier-Stokes equations do not disclose any
other distinguishing length scales, thus suggesting a picture of a
scale-invariant inertial-range dynamics \cite{MON71,FRI95,
comment}. This scale-invariance should reflect itself in the
scaling of structure functions, which are moments of velocity
increments constructed from the measured velocity data. However,
measured structure functions in shear flows seem to show only an
approximate multiscaling character, even at the largest accessible
Reynolds numbers, where the large-scale and the dissipation scale
are separated by five orders of magnitude
\cite{ARN96,SRE97,SREE98}. The reason for this distortion of
scaling seems to be the sensitivity of structure functions to the
mean shear that is inevitably present in most natural flows at
high Reynolds numbers. While schemes have been proposed to account
for these effects \cite{arad,SREE98}, it has been demonstrated
recently \cite{CLE03} that the lowest-order two-point correlation
function of the surrogate energy dissipation of a
high-Reynolds-number atmospheric boundary layer reveals a far more
convincing power-law scaling over the entire inertial range,
without having to resort to shear corrections of the sort needed
for structure functions. This finding suggests that the
phenomenological picture of the self-similar turbulent energy
cascade may contain more truth than previously anticipated.
Consequently, it is reasonable to reexamine simple empirical
energy cascade models and extend to a higher level the comparison
between the models and experimental data. This is the purpose of
the paper.

Random multiplicative cascade processes (RMCP) present a
particularly simple geometrical picture of the energy cascade
\cite{FRI95} and are a natural archetype for multiscaling. In
their binary version, for example \cite{MS}, they introduce a
hierarchy of length scales $l_j=L/2^j$ and a random multiplicative
cascade generator is used to transport the energy flux from the
integral scale $L$ through the inertial range scales $L\geq
l_j\geq\eta$ down to the dissipation scale $\eta=L/2^J$. Referring
to the ultrametric branching structure of binary RMCPs, $N$-point 
correlation functions of arbitrary order have been calculated 
analytically with generating function techniques
\cite{GRE95,GRE96,GRE98a,GRE98b}. In terms of the ultrametric
two-point distance, which measures the number of cascade steps
necessary to reach the last common branching, these models reveal 
rigorous multiscaling. However, in this form, the two-point RMCP 
statistics cannot be compared directly to quantities that can be 
extracted from the data because, from an experimentalist's 
perspective, the two-point correlations are expressed as functions 
of the Euclidean two-point distance. In this respect, RMCPs are 
incomplete and have to be supplemented by model-dependent extensions, 
stating the conversion of the ultrametric to Euclidean statistics
\cite{JOU02,EGG01,MEN90}. It may be expected that this unavoidable
conversion will lead to some deviations from rigorous multiscaling
of correlation functions for two-point distances within the
inertial range. The challenge is to find the degree to which the
theoretical, extended-RMCP, two-point correlation functions can
match their experimental counterparts.

If this comparison is satisfactory, the two-point correlation
statistics of the energy dissipation will allow for an unambiguous
extraction of multiscaling exponents, these being related to
moments of the RMCP generator. In addition, two-point cumulants of
logarithmic energy dissipation \cite{EGG01} allow us to access
properties of the cascade generator in another way. Taken
together, reliable parametric estimations of the underlying
cascade generator can then be given. This would be the best we can
do to achieve a data-driven derivation of the turbulent energy
cascade generator and to settle the question of whether the latter
is bimodal \cite{SCH85,MEN87}, log-normal \cite{OBU62,KOL62},
log-stable \cite{KID91,SCH92}, log-infinite divisible
\cite{SHE94,DUB94,SHE95} or another type of distribution. None of
those prototype cascade generators has been directly derived from
data.

Note that the breakup coefficients, sometimes called multipliers,
have once been thought to represent a direct approach to derive
the RMCP cascade generator from data
\cite{CHAB92,SRE95,MOL95,PED96,NEL96}. In fact, generator-like
scale-invariant distributions of breakup coefficients have been
observed, but conditional distributions have been found to exhibit
scale correlations. In a series of papers \cite{JOU99,JOU00,JOU02}
these findings have been fully explained within ultrametric
hierarchical RMCPs, once those are analyzed from an
experimentalist's perspective, including unavoidable small-scale
resummation and restoration of spatial homogeneity via the
ultrametric-Euclidean conversion. This work has demonstrated that
cascade generators and distributions of breakup coefficients are
not directly related.

The structure of this paper is as follows. Section II provides
basic information about the two-point correlation of the lowest
order for the surrogate energy dissipation in four fully developed
turbulent flows---one atmospheric boundary layer and three
wind-tunnel flows. This information serves to provide the needed
guidance for the subsequent RMCP modelling of Sec.\ III, where
analytic expressions for two-point correlators and cumulants are
derived, taking into account the ultrametric-Euclidean conversion.
More information on data analysis is provided in Sec.\ IV, leading
in Sec.\ V to further comparisons between prototype distributions
and experiments. As will also be demonstrated in this section, the
extracted properties of the generator allow us to make parametric
estimations of the RMCP cascade generator. A conclusion and
outlook is given in Sec.\ VI.

%%%%%%%%%%%%%%%%%%%%%%%%%%%%%%%%%%%%%%%%%%%%%%%%%%%%%%%%%%%%%%%%%%%%%%%
%%%%%%%%%%%%%%%%%%%%%%%%%%%%%%%%%%%%%%%%%%%%%%%%%%%%%%%%%%%%%%%%%%%%%%%
\section{Data analysis I: basic facts on two-point statistics}
%%%%%%%%%%%%%%%%%%%%%%%%%%%%%%%%%%%%%%%%%%%%%%%%%%%%%%%%%%%%%%%%%%%%%%%
%%%%%%%%%%%%%%%%%%%%%%%%%%%%%%%%%%%%%%%%%%%%%%%%%%%%%%%%%%%%%%%%%%%%%%%

We analyze four different data sets, three of which have been
recorded in a wind tunnel \cite{PEA02} and the fourth in an
atmospheric boundary layer \cite{DHR00} about 35 m above the
ground. We will refer to them as w1, w2, w3 and a, respectively.
Characteristic quantities of all data sets are summarized in Table
\ref{table1}. The Reynolds number $R_\lambda = \sqrt{\langle
u^2\rangle} \lambda/\nu$ is based on the Taylor microscale
$\lambda = \sqrt{ \langle u^2 \rangle / \langle( \partial
u/\partial x)^2 \rangle}$; $\nu$ is the kinematic viscosity and
$u$ is the streamwise velocity component. Upon the application of
the frozen flow hypothesis, the recorded time series were
converted into one-dimensional spatial series. The energy spectra
of all four records reveal more or less the typical $5/3$ slope in
the inertial range. In contrast to wind-tunnel records, that from
the atmospheric boundary layer reveals a white-noise behavior at
very small scales; this noise, which comes from detailed
electronic circuitry, has been removed by an appropriate Wiener
filter. The energy dissipation was then calculated as the
surrogate amplitude $\varepsilon = 15 \nu (\frac{du}{dx})^2$.
Various tests were made to ensure that the effect of Wiener
filtering were not consequential.

Figure 1 illustrates the lowest-order two-point correlator
$r_{1,1}(d=|x_2-x_1|) =
 \langle \varepsilon(x_1) \varepsilon(x_2) \rangle
 / ( \langle \varepsilon(x_1) \rangle
     \langle \varepsilon(x_2) \rangle )$
sampled from the four different experimental records. Well inside
the inertial range $\eta \ll d \ll L$ the two-point correlators
reveal a power-law behavior $r_{1,1} \sim (L/d)^{\tau_{1,1}}$.
Power-law fits are indicated by the shifted broken straight lines;
see also the insets, where the local slopes are shown. The
resulting scaling exponents are $\tau_{1,1}=0.15$ (w1), $0.14$
(w2), $0.18$ (w3) and $0.20$ (a). Note that there is a Reynolds
number dependence of this exponent. This has been explored in
greater detail elsewhere \cite{CLE03a}.

For the records with the largest Reynolds number, there is a large
scale range for which the two-point correlator exhibits a rigorous
power-law scaling. However, with decreasing Reynolds number this
scale range becomes smaller. As a rule of thumb, we observe that a
good scaling range is confined between $d \approx 20{-}30 \eta$
and $\approx 0.5 L$. If we could understand precisely the 
deviations from the power-law scaling beyond this intermediate
inertial range, a more satisfactory extraction of multiscaling
exponents would be possible, especially for turbulent flows with
moderate Reynolds numbers. In the small-distance 
$d \leq 20{-}30 \eta$ region, however, the correlations observed 
exceed the power-law extrapolation; consult Fig.\ 1 again. As has 
been explained in Ref.\ \cite{CLE03}, this enhancement is a 
consequence of the unavoidable surrogacy of the experimentally 
measured energy dissipation. Without knowing the correct 
small-distance behavior based on the proper energy dissipation, a 
theoretical modelling of two-point correlations in the dissipative 
regime is not very meaningful. We are thus left to inspect only the 
remainder of the inertial range. Fortunately, this range is 
sufficiently large, especially for the atmospheric data.

At $d \approx L$, the two-point correlator has not yet converged
to unity, which is its asymptotic value as $d{\to}\infty$. For all
four data records inspected, the decorrelation length appears to
be around $L_\mathrm{dec} \approx 4 L$ and matches the length
observed in the autocorrelation function of the streamwise
velocity component. These findings suggest that the two-point
correlator can be described as
%----------------------------------------------------------------------
\begin{equation}
\label{two1}
  r_{1,1}(20{-}30 \eta < d \leq L_\mathrm{dec})
    =  a_{1,1}
       \left( \frac{L_\mathrm{dec}}{d} \right)^{\tau_{1,1}}
       f_{1,1}(d/L_\mathrm{dec})
       \; .
\end{equation}
%----------------------------------------------------------------------
For two-point distances $d$ much smaller than the decorrelation
length the finite-size function converges to unity; that is,
$f_{1,1}(d \ll L_\mathrm{dec}) \to 1$. One goal of this paper is
to qualitatively and quantitatively reproduce this functional form
with an extended modelling based on random multiplicative cascade
processes. This approach, which is the subject of the next
section, naturally suggests a physical interpretation of the
decorrelation length $L_\mathrm{dec}$.

%%%%%%%%%%%%%%%%%%%%%%%%%%%%%%%%%%%%%%%%%%%%%%%%%%%%%%%%%%%%%%%%%%%%%%%
%%%%%%%%%%%%%%%%%%%%%%%%%%%%%%%%%%%%%%%%%%%%%%%%%%%%%%%%%%%%%%%%%%%%%%%
\section{Two-point statistics of random multiplicative cascade
         processes}
%%%%%%%%%%%%%%%%%%%%%%%%%%%%%%%%%%%%%%%%%%%%%%%%%%%%%%%%%%%%%%%%%%%%%%%
%%%%%%%%%%%%%%%%%%%%%%%%%%%%%%%%%%%%%%%%%%%%%%%%%%%%%%%%%%%%%%%%%%%%%%%

%%%%%%%%%%%%%%%%%%%%%%%%%%%%%%%%%%%%%%%%%%%%%%%%%%%%%%%%%%%%%%%%%%%%%%%
\subsection{Binary random multiplicative cascade process}
%%%%%%%%%%%%%%%%%%%%%%%%%%%%%%%%%%%%%%%%%%%%%%%%%%%%%%%%%%%%%%%%%%%%%%%

In its simplest form an RMCP employs a binary hierarchy of length
scales $l_j=L_\mathrm{casc}/2^j$. In the first cascade step the
parent interval of length $l_0 = L_\mathrm{casc}$ is split into
left and right daughter intervals, both of length $l_1$. In
subsequent cascade steps, each interval of generation $0 \leq j
\leq J-1$ is again split into a left and right subinterval of
length $l_{j+1}=l_j/2$. Once the dissipation scale
$\eta=L_\mathrm{casc}/2^J$ is reached, the interval splitting
stops and has resulted into $2^J$ spatially ordered intervals of
smallest size $\eta$. It is convenient to label them as well as
their ancestors according to the binary notation
$\mathbf{\kappa}^{(j)}=\kappa_1,\kappa_2,\ldots,\kappa_j$. The
label refers to the hierarchical position of an interval of
generation $j$, where $\kappa_i=0$ or $1$ stands for the left or
right interval, respectively.

The binary interval splittings go together with a probabilistic
evolution of the energy-flux field. From generation $j$ to $j+1$
the field amplitudes propagate locally as
%----------------------------------------------------------------------
\begin{eqnarray}
\label{threea1}
  \Pi_{\mathbf{\kappa}^{(j)},0}
    &=&  q_{\mathbf{\kappa}^{(j)},0} \Pi_{\mathbf{\kappa}^{(j)}}
         \; , \nonumber \\
  \Pi_{\mathbf{\kappa}^{(j)},1}
    &=&  q_{\mathbf{\kappa}^{(j)},1} \Pi_{\mathbf{\kappa}^{(j)}}
         \; .
\end{eqnarray}
%----------------------------------------------------------------------
The two random multiplicative weights $q_\mathrm{left} =
q_{\mathbf{\kappa}^{(j)},0}$ and $q_\mathrm{right} =
q_{\mathbf{\kappa}^{(j)},1}$, with mean $\langle q_\mathrm{left}
\rangle = \langle q_\mathrm{right} \rangle =1$, are drawn from a
scale-independent bivariate probability density function
$p(q_\mathrm{left},q_\mathrm{right})$, which is called the cascade
generator. Initially, corresponding to $j=0$, the iteration
(\ref{threea1}) starts with a given large-scale energy flux $\Pi$,
which might itself be a random variable fluctuating around its
normalized mean $\langle \Pi \rangle = 1$. After the last
iteration $J-1 \to J$, the energy-flux amplitude is interpreted as
the amplitude $\varepsilon_{\mathbf{\kappa}^{(J)}} =
\Pi_{\mathbf{\kappa}^{(J)}}$ of the energy dissipation supported
at the interval of length $\eta$ at position
$\mathbf{\kappa}^{(J)}$. As a result of (\ref{threea1}), this
amplitude is a multiplicative sum of the random weights, given by
%----------------------------------------------------------------------
\begin{equation}
\label{threea2}
  \varepsilon_{\kappa_1,\ldots,\kappa_J}
    =  q_{\kappa_1,\ldots,\kappa_J}
       q_{\kappa_1,\ldots,\kappa_{J-1}}
       \cdots q_{\kappa_1} \Pi
       \; .
\end{equation}
%----------------------------------------------------------------------

In the following we assume the cascade generator to be of the
factorized form $p(q_\mathrm{left},q_\mathrm{right}) =
p(q_\mathrm{left}) p(q_\mathrm{right})$ with identical statistics
for the left/right variable. Of course, the factorization is not
the most general ansatz but, as already pointed out in Ref.\
\cite{JOU00}, it represents a reasonable approximation: the
turbulent energy cascade takes place in three spatial dimensions
and calls for a three-dimensional RMCP modelling, respecting
energy conservation. Since the measured temporal records come from
one-dimensional cuts, the three-dimensional RMCP has to be observed 
in unity subdimension. Because of this, the RMCP appears to be 
non-conservative and the two multiplicative weights appear to be 
almost decorrelated and independent of each other.

%%%%%%%%%%%%%%%%%%%%%%%%%%%%%%%%%%%%%%%%%%%%%%%%%%%%%%%%%%%%%%%%%%%%%%%
\subsection{Two-point correlators}
%%%%%%%%%%%%%%%%%%%%%%%%%%%%%%%%%%%%%%%%%%%%%%%%%%%%%%%%%%%%%%%%%%%%%%%

Expressions for $N$-point moments $\langle \varepsilon(x_1)^{n_1}
\cdots \varepsilon(x_N)^{n_N} \rangle$ are easily found. They can
be calculated either by a straightforward approach or, more
formally, by an iterative construction of the respective
multivariate characteristic function \cite{GRE95,GRE96}; a third
and more elegant approach \cite{BIA99} makes use of the full
analytic solution of the multivariate characteristic function for
logarithmic cascade-field amplitudes \cite{GRE98a,GRE98b}. We
simply state the results up to two-point correlations:
%----------------------------------------------------------------------
\begin{eqnarray}
\label{twob1}
  \left\langle \varepsilon(x_1)^{n_1} \right\rangle
    &=&  \left\langle q^{n_1} \right\rangle^J
         \left\langle \Pi^{n_1} \right\rangle
         \; , \\
\label{twob2}
  \left\langle
  \varepsilon(x_1)^{n_1}
  \varepsilon(x_2)^{n_2}
  \right\rangle
    &=&  \left\langle q^{n_1+n_2} \right\rangle^{J-D}
         \left\langle q^{n_1} \right\rangle^{D}
         \left\langle q^{n_2} \right\rangle^{D}
         \left\langle \Pi^{n_1+n_2} \right\rangle
         \; .
\end{eqnarray}
%----------------------------------------------------------------------
Here, the binary notation $\mathbf{\kappa}^{(J)}$ has been
transformed into a spatial bin label
$x=1+\sum_{j=1}^J\kappa_j2^{J-j}$, which runs over $1\leq x \leq
2^J$ in units of $\eta$. Two bins $x_1 \equiv
(\kappa_1,\ldots,\kappa_{J-D},\kappa_{J-D+1},\ldots,\kappa_J)$ and
$x_2 \equiv
(\kappa_1,\ldots,\kappa_{J-D},\kappa^\prime_{J-D+1},\ldots,
\kappa^\prime_J)$ are assigned an ultrametric distance $D$ once
the first $J-D$ $\kappa$'s are identical and $\kappa_{J-D+1} \neq
\kappa^\prime_{J-D+1}$. In other words, after $J-D$ common
branches along the binary tree, the two bins separate into
different branches.

For the extraction of scaling exponents
%----------------------------------------------------------------------
\begin{equation}
\label{twob4}
  \tau_n
    =  \log_2\langle{q^n}\rangle
       \; ,
\end{equation}
%----------------------------------------------------------------------
it is enough to consider the two-point statistics (\ref{twob2}).
In normalized form, the two-point correlators are found to scale 
perfectly as
%----------------------------------------------------------------------
\begin{eqnarray}
\label{twob5}
  r_{n_1,n_2}(D)
    &=&  \frac{ \left\langle
                \varepsilon(x_1)^{n_1}
                \varepsilon(x_2)^{n_2}
                \right\rangle }
              { \left\langle
                \varepsilon(x_1)^{n_1}
                \right\rangle \left\langle
                \varepsilon(x_2)^{n_2}
                \right\rangle }
         \nonumber \\
    &=&  \frac{ \langle \Pi^{n_1+n_2} \rangle }
              { \langle \Pi^{n_1} \rangle
                \langle \Pi^{n_2} \rangle }
         \left(
         \frac{ \left\langle q^{n_1+n_2} \right\rangle }
              { \left\langle q^{n_1} \right\rangle
                \left\langle q^{n_2} \right\rangle }
         \right)^{J-D}
         \nonumber \\
    &=&  \frac{ \langle \Pi^{n_1+n_2} \rangle }
              { \langle \Pi^{n_1} \rangle
                \langle \Pi^{n_2} \rangle }
         \left(
         \frac{L_{\rm casc}}{l_D}
         \right)^{\tau_{n_1,n_2}}
         \; ,
\end{eqnarray}
%----------------------------------------------------------------------
where $l_D=2^{D-1}\eta$ represents the characteristic two-bin
distance corresponding to the ultrametric distance $D>0$ and
%----------------------------------------------------------------------
\begin{equation}
\label{twob6}
  \tau_{n_1,n_2}
    =  \tau_{n_1+n_2} - \tau_{n_1} - \tau_{n_2}
       \; .
\end{equation}
%----------------------------------------------------------------------

For an experimentalist, the expression (\ref{twob5}) does not
present an observable result. Different pairs of bins, all having
an identical Euclidean distance $\eta{\leq}d{<}L_{\rm casc}$, do
not have an unequivocal ultrametric distance. Depending on their
position within the binary ultrametric cascade tree, the two bins
might share a cascade history that is long (small $D$) or short
(large $D$). Consequently, as an experimentalist analyzes the
two-point statistics in terms of $d$, the ultrametric expression
(\ref{twob5}) has to be averaged over all $D$ that contribute to
the same value of $d$. In order to perform this conversion from an
ultrametric to an Euclidean distance and, by this means, to
restore spatial homogeneity, we introduce the discrete conditional
probability distribution
%----------------------------------------------------------------------
\begin{equation}
\label{twob7}
  p(D|d)
    =  \left\{
       \begin{array}{lcl}
       0            & \qquad &
                        (1{\leq}D{<}A{=}\lceil\log_2d\rceil)  \\
       1 - d2^{-D}  &&  (D{=}A)  \\
       d2^{-D}      &&  (A{<}D{\leq}J)  \\
       0            &&  (J{<}D{<}\infty)  \\
       d2^{-J}      &&  (D{=}\infty)
       \end{array}
       \right.
\end{equation}
%----------------------------------------------------------------------
of finding the ultrametric distance $D$ for a given Euclidean
distance $d$ in units of $\eta$ \cite{EGG01}. This expression has
been derived by employing the chain picture of independent cascade
configurations; consult Fig.\ 2. It roughly goes as $p(D|d) \sim
2^{\log_2{d}-D}$. The sum $\sum_{D=0}^J p(D|d) = 1 - d2^{-J}$ does
not add up to unity, since $p(\infty|d) = d2^{-J}$ represents the
probability that the two bins belong to different $L_{\rm
casc}$-domains.

Since the one-point statistics $\langle\varepsilon(x)^{n}\rangle =
\langle\varepsilon(x+d)^{n}\rangle$ do not depend on the spatial
index $x$, the ultrametric-Euclidean conversion of the normalized
two-point correlator (\ref{twob5}) leads to
%----------------------------------------------------------------------
\begin{eqnarray}
\label{twob8}
  r_{n_1,n_2}(d{\neq}0)
    &=&  \sum_{D=1}^J p(D|d) r_{n_1,n_2}(D)
         + p(\infty|d)
         \nonumber \\
    &=&  \frac{\langle\Pi^{n_1+n_2}\rangle}
              {\langle\Pi^{n_1}\rangle
               \langle\Pi^{n_2}\rangle}
         \left(
         1 - \frac{d}{2^A}
         + \frac{d}{2^A}
         \left( 2\frac{\langle{q^{n_1+n_2}}\rangle}
                      {\langle{q^{n_1}}\rangle\langle{q^{n_2}}\rangle}
                - 1 \right)^{-1}
         \right)
         \left(
         \frac{ \langle q^{n_1+n_2} \rangle }
              { \langle q^{n_1} \rangle \langle q^{n_2} \rangle }
         \right)^{J-A}
         \nonumber \\
    & &  + \left(
         1 -
         \frac{\langle\Pi^{n_1+n_2}\rangle}
              {\langle\Pi^{n_1}\rangle
               \langle\Pi^{n_2}\rangle}
         \left( 2\frac{\langle{q^{n_1+n_2}}\rangle}
                      {\langle{q^{n_1}}\rangle\langle{q^{n_2}}\rangle}
                - 1 \right)^{-1}
         \right)
         \frac{d}{2^J}
         \; .
\end{eqnarray}
%----------------------------------------------------------------------
This expression holds for every $\eta{\leq}d{\leq}L_{\rm casc}$. For
$d{=}0$ and $d{>}L_{\rm casc}$ the normalized two-point density simply
becomes
$r_{n_1,n_2}(d{=}0)
 = \langle\Pi^{n_1+n_2}\rangle\langle q^{n_1+n_2}\rangle^J
   / ( \langle\Pi^{n_1}\rangle\langle q^{n_1}\rangle^J
       \langle\Pi^{n_2}\rangle\langle q^{n_2}\rangle^J ) $
and $r_{n_1,n_2}(d{>}L_{\rm casc})=1$, respectively. The two-point
density (\ref{twob8}) does not reveal perfect scaling anymore.
Usually the second term, scaling as $d/L_{\rm casc}$, is small
when compared to the first term, except for $d{\approx}L_{\rm
casc}$. The modulations, observed for the first term, are an
artifact of the discrete scale invariance \cite{SOR98} of the
binary random multiplicative cascade model implementation. In the
following we will discard these modulations by first considering
only dyadic distances $d_m=L_{\rm casc}/2^m$ with integer
$0{\leq}m{<}J$, and then switching again to continuous $d$ by
interpolating between the discrete $d_m$. The expression
(\ref{twob8}) then simplifies to
%----------------------------------------------------------------------
\begin{equation}
\label{twob9}
  r_{n_1,n_2}(d)
    =  a_{n_1,n_2}
       \left( \frac{L_{\rm casc}}{d} \right)^{\tau_{n_1,n_2}}
       f_{n_1,n_2}(d/L_{\rm casc}),
\end{equation}
%----------------------------------------------------------------------
with
%----------------------------------------------------------------------
\begin{equation}
\label{twob10}
  a_{n_1,n_2}
    =  \frac{\langle\Pi^{n_1+n_2}\rangle}
            {\langle\Pi^{n_1}\rangle
             \langle\Pi^{n_2}\rangle}
       \left( 2\frac{\langle{q^{n_1+n_2}}\rangle}
                    {\langle{q^{n_1}}\rangle\langle{q^{n_2}}\rangle}
              - 1 \right)^{-1}
\end{equation}
%----------------------------------------------------------------------
and the finite-size scaling function
%----------------------------------------------------------------------
\begin{equation}
\label{twob11}
  f_{n_1,n_2}(d/L_{\rm casc})
    =  1 +
       \left(
       a_{n_1,n_2}^{-1}
       - 1
       \right)
       \left( \frac{d}{L_{\rm casc}} \right)^{1+\tau_{n_1,n_2}}
       \; .
\end{equation}
%----------------------------------------------------------------------
Figure 3 compares the expressions (\ref{twob8}) and (\ref{twob9})
for the order $n_1=n_2=1$.

The finite-size scaling function has the property
$f_{n_1,n_2}(d{\ll}L_{\rm casc}) = 1$ as long as the condition
$1+\tau_{n_1,n_2} \geq 0$ or, equivalently, $\langle q^{n_1+n_2}
\rangle / (\langle q^{n_1} \rangle \langle q^{n_2} \rangle) > 1/2$
is fulfilled. This is the case for all positive combinations
$n_1\geq 0,n_2\geq 0$. However, combinations with negative orders
do exist, for which the second term on the right hand side
of (\ref{twob11}) then dominates over the first term in the limit
$d/L_{\rm casc}\to 0$. This implies that the normalized two-point
density (\ref{twob9}) asymptotically scales as
$r_{n_1,n_2}(d)\sim(L_{\rm casc}/d)^{-1}$, giving rise to the
effective scaling exponents $\tau^{\rm
eff}_{n_1,n_2}=\sup\{-1,\tau_{n_1,n_2}\}$. This scaling transition
is again a pure consequence of the ultrametric-Euclidean
conversion. More discussions on this subject can be found in
Refs.\ \cite{MEN90,ONE93,BEN97}.

Upon studying the expression (\ref{twob11}) more closely, we
realize that two effects, ultrametric-Euclidean conversion and
large-scale fluctuations, contribute to the finite-size scaling
function. They have a tendency to cancel each other. Once we have
%----------------------------------------------------------------------
\begin{equation}
\label{twob12}
  \frac{ \langle \Pi^{n_1+n_2} \rangle }
       { \langle \Pi^{n_1} \rangle
         \langle \Pi^{n_2} \rangle }
    =  2 \frac{ \langle q^{n_1+n_2} \rangle }
              { \langle q^{n_1} \rangle \langle q^{n_2} \rangle }
       - 1
       \; ,
\end{equation}
%----------------------------------------------------------------------
the finite-size scaling function becomes $f_{n_1,n_2}(d/L_{\rm
casc})=1$ exactly, showing no $d$-dependence.

We also wish to point out an interesting mathematical observation
following from the specific expressions
(\ref{twob9})-(\ref{twob11}). Since the finite-size scaling function
(\ref{twob11}) reveals the simple scaling behavior
%----------------------------------------------------------------------
\begin{equation}
\label{twob13}
  \left( f_{n_1,n_2}(d/L_{\rm casc}) - 1 \right)
  \left( \frac{L_{\rm casc}}{d} \right)^{\tau_{n_1,n_2}}
    \sim  \left( \frac{d}{L_{\rm casc}} \right)^\mu
\end{equation}
%----------------------------------------------------------------------
with $\mu=1$, we find
%----------------------------------------------------------------------
\begin{equation}
\label{twob14}
  r_{n_1,n_2}(d)
  - \frac{1}{\xi^\mu} r_{n_1,n_2}(\xi d)
    =  a_{n_1,n_2}
       \left( 1 - \frac{1}{\xi^{\mu+\tau_{n_1,n_2}}} \right)
       \left( \frac{L_{\rm casc}}{d} \right)^{\tau_{n_1,n_2}}
       \; ,
\end{equation}
%----------------------------------------------------------------------
where the normalized two-point correlator with the rescaled two-point 
distance $\xi d$ has been subtracted from itself. As a function of the
two-point distance $d$ this quantity exhibits a rigorous power-law 
behavior with scaling exponents $\tau_{n_1,n_2}$, which is independent 
of the chosen rescaling parameter $\xi$ and is free of large-scale 
effects.

%%%%%%%%%%%%%%%%%%%%%%%%%%%%%%%%%%%%%%%%%%%%%%%%%%%%%%%%%%%%%%%%%%%%%%%
\subsection{Two-point cumulants}
%%%%%%%%%%%%%%%%%%%%%%%%%%%%%%%%%%%%%%%%%%%%%%%%%%%%%%%%%%%%%%%%%%%%%%%

Because experimental data yield limited statistics, the two-point
correlation densities (\ref{twob9}) will be restricted to lowest
orders $1\leq n_1+n_2\leq 3$ or $4$. This limits us to indirect
information on the cascade generator, namely the scaling exponents
$\tau_1, \tau_2, \tau_3$ and, perhaps, $ \tau_4$ of (\ref{twob4}).
In order to do better, we need to accumulate additional and
complementary information. In fact, as proposed already in Ref.\
\cite{EGG01}, this can be achieved by switching to the logarithmic
amplitude $\varepsilon(x)\to\ln\varepsilon(x)$, and from two-point
correlation densities to two-point cumulants
%%%---------------------------------------------------------------
\begin{eqnarray}
\label{twoc1}
  C_{1,1}(x_2{-}x_1)
    &=&  \left\langle
         \ln\varepsilon(x_1) \ln\varepsilon(x_2)
         \right\rangle
         - \left\langle \ln\varepsilon(x) \right\rangle^2
         \; ,  \nonumber \\
  C_{2,1}(x_2{-}x_1)
    &=&  \left\langle
         \ln^2\varepsilon(x_1) \ln\varepsilon(x_2)
         \right\rangle
         - 2\left\langle
           \ln\varepsilon(x_1) \ln\varepsilon(x_2)
           \right\rangle
           \left\langle \ln\varepsilon(x) \right\rangle
         - \left\langle \ln^2\varepsilon(x) \right\rangle
           \left\langle \ln\varepsilon(x) \right\rangle
         + 2 \left\langle \ln\varepsilon(x) \right\rangle^3
         \; , \ldots , \nonumber \\
  C_{n_1,n_2}(x_2{-}x_1)
    &=&  \left.
         \frac{\partial^{n_1+n_2}}
              {\partial\lambda^{n_1}\partial\lambda^{n_2}}
         \ln\left\langle
         \varepsilon(x_1)^{\lambda_1} \varepsilon(x_2)^{\lambda_2}
         \right\rangle
         \right|_{\lambda_1=\lambda_2=0}
         \; .
\end{eqnarray}
%%%---------------------------------------------------------------
Explicit RMCP expressions have already been derived in Ref.\
\cite{EGG01} within the ultrametric view as well as the converted
ultrametric-Euclidean view. Here, we summarize only the latter
lowest-order results, which hold for $\eta\leq d\leq L_{\rm
casc}$:
%%%---------------------------------------------------------------
\begin{equation}
\label{twoc2}
  C_{n-1,1}(d)
    =  G_1(J,d) c_n
       +
       G_0(J,d) \langle \ln^n\Pi \rangle_c
       \; .
\end{equation}
%%%---------------------------------------------------------------
The geometric functions $G_n(J,d) = (1/2^J) \sum_{D=1}^J (J-D)^n
p(D|d)$ are related to moments of the conditional probability
distribution (\ref{twob7}) and are fingerprints of the
hierarchical RMCP tree structure. They are given by the
expressions
%%%---------------------------------------------------------------
\begin{eqnarray}
\label{twoc3}
  G_0(J,d)
    &=&  1 - \frac{d}{L_{\rm casc}}
         \; ,  \nonumber \\
  G_1(J,d)
    &=&  (J-A) - 2\frac{d}{\eta} \left( 2^{-A}-2^{-J} \right)
         \; \approx \;
         \log_2\left( \frac{L_{\rm casc}}{d} \right)
         - 2 + 2\frac{d}{L_{\rm casc}}
         \; ,
\end{eqnarray}
%%%---------------------------------------------------------------
with the last step neglecting small log-oscillations. The
cumulants of the logarithmic multiplicative weight
%%%---------------------------------------------------------------
\begin{equation}
\label{twoc4}
  c_{n}
    =  \left\langle \ln^n q \right\rangle_c
    =  \left.
       \frac{\partial^{n} Q(\lambda)}{\partial \lambda^{n}}
       \right|_{\lambda=0}
\end{equation}
%%%---------------------------------------------------------------
are generated by the logarithm of the Mellin transform of the
cascade generator, i.e.,
%%%---------------------------------------------------------------
\begin{equation}
\label{twoc5}
  Q(\lambda)
    =  \ln\left( \int dq p(q) q^\lambda \right)
       \; .
\end{equation}
%%%---------------------------------------------------------------
The cumulants $\langle\ln^n\Pi\rangle_c$ of the initial large-scale
fluctuation are defined analogous to $\langle\ln^n q\rangle_c$.

%%%%%%%%%%%%%%%%%%%%%%%%%%%%%%%%%%%%%%%%%%%%%%%%%%%%%%%%%%%%%%%%%%%%%%%
\subsection{Multifractal sum rules}
%%%%%%%%%%%%%%%%%%%%%%%%%%%%%%%%%%%%%%%%%%%%%%%%%%%%%%%%%%%%%%%%%%%%%%%

The cumulants $c_n$ of (\ref{twoc4}) and the scaling
exponents $\tau_n$ of (\ref{twob4}) are not independent of each
other. Combining Eqs.\ (\ref{twob4}), (\ref{twoc4}) and
(\ref{twoc5}), we arrive at
%%%---------------------------------------------------------------
\begin{equation}
\label{twoc6}
  Q(n)
    =  \ln 2 \; \tau_n
    =  \sum_{k=1}^\infty c_k \frac{n^k}{k!}
       \; .
\end{equation}
%%%---------------------------------------------------------------
In the lowest order, this translates to
%%%---------------------------------------------------------------
\begin{equation}
\label{twoc7}
  \ln 2 \; \tau_1 = 0
    =  c_1 + \frac{c_2}{2} + \frac{c_3}{6} + \ldots
       \; .
\end{equation}
%%%---------------------------------------------------------------
These multifractal sum rules can be used, for example, to estimate
$c_1$, which can not be extracted from the two-point statistics
(\ref{twoc2}).

Another approach to estimate the value of $c_1$ is given by the
well-known replica trick:
%%%---------------------------------------------------------------
\begin{equation}
\label{twoc8}
  c_1
    =  \langle \ln q \rangle
    =  \left.
       \frac{\partial Q(\lambda)}{\partial\lambda}
       \right|_{\lambda=0}
    =  \left.
       \frac{\partial \left\langle q^\lambda \right\rangle}
            {\partial\lambda}
       \right|_{\lambda=0}
    =  \lim_{\lambda\to 0}
       \frac{\langle q^\lambda \rangle - 1}{\lambda}
    =  \lim_{\lambda\to 0}
       \frac{2^{\tau_\lambda}-1}{\lambda}
       \; .
\end{equation}
%%%---------------------------------------------------------------
Another form of sum rules follows from (\ref{twoc6}) and states that
%%%---------------------------------------------------------------
\begin{equation}
\label{twoc9}
  \left.
  \ln 2 \, \frac{\partial\tau_\lambda}{\partial\lambda}
  \right|_{\lambda=n}
    =  \sum_{k=0}^\infty c_{k+1} \frac{n^k}{k!}
    =  c_1 + n c_2 + \frac{n^2}{2} c_3 + \ldots
       \; .
\end{equation}
%%%---------------------------------------------------------------
It can be seen as a generalization of (\ref{twoc8}).

%%%%%%%%%%%%%%%%%%%%%%%%%%%%%%%%%%%%%%%%%%%%%%%%%%%%%%%%%%%%%%%%%%%%%%%
%%%%%%%%%%%%%%%%%%%%%%%%%%%%%%%%%%%%%%%%%%%%%%%%%%%%%%%%%%%%%%%%%%%%%%%
\section{Data analysis II: more on two-point statistics}
%%%%%%%%%%%%%%%%%%%%%%%%%%%%%%%%%%%%%%%%%%%%%%%%%%%%%%%%%%%%%%%%%%%%%%%
%%%%%%%%%%%%%%%%%%%%%%%%%%%%%%%%%%%%%%%%%%%%%%%%%%%%%%%%%%%%%%%%%%%%%%%

In this section, we provide further aspects of data analysis that 
are relevant. The goal is three-fold: first and already pointed out
in Sect.\ II, to test the RMCP expression (\ref{twob9}) with the
proposed finite-size scaling for two-point correlators, second, to
test the expression (\ref{twoc2}) for two-point cumulants derived 
from the RMCP theory, and, third, to extract reliable values for 
the scaling exponents $\tau_n$ and cumulants $c_n$ from various 
turbulent records discussed in Sec.\ II. 

The expressions (\ref{twob9}) and (\ref{twoc2}) come with
parameters $L_\mathrm{casc}$, $\tau_{n_1,n_2}$, $a_{n_1,n_2}$,
$c_n$ and $\langle \ln^n\Pi \rangle_c$. The parameter
$L_\mathrm{casc}$ depends on neither the order $n_1,n_2$ nor the
choice of the two-point statistics, i.e., whether we use the
correlator or the cumulant. The fits of (\ref{twob9}) and
(\ref{twoc2}) to their counterparts from experimental records have
to respect this independence. In addition to the common parameter
$L_\mathrm{casc}$, the fit of each order has two more parameters:
these are $\tau_{n_1,n_2}$, $a_{n_1,n_2}$ for the two-point
correlator and $c_n$, $\langle \ln^n\Pi \rangle_c$ for the two-point
cumulant. We demonstrate the quality of fits by choosing the data
record w2. For this data set, two-point correlators with orders
from $n_1,n_2=1,1$ to $2,2$ are illustrated in Fig.\ 4, while the
two-point cumulants with orders $n_1,n_2=1,1$ and $2,1$ are
illustrated in Fig.\ 5. Except for very small two-point distances
$\eta\leq d\leq 20{-}30\eta$, where, as already noted, the
surrogacy effect of the energy dissipation corrupts the
experimental two-point statistics \cite{CLE03}, the agreement
between the experimental two-point correlators and cumulants and
the best-fit expressions (\ref{twob9}) and (\ref{twoc2}) is
remarkable. Table II lists the best-fit parameters
$L_\mathrm{casc}$, $\tau_n$ and $c_n$. Note that, due to
(\ref{twob6}) and $\tau_1=0$, the scaling exponents
$\tau_{n_1,n_2}$ have been converted into $\tau_2=\tau_{1,1}$,
$\tau_3=\tau_{2,1}+\tau_2$ and
$\tau_4=\tau_{3,1}+\tau_3=\tau_{2,2}+2\tau_2$. For w2 the two
values $\tau_4=0.79$ and $0.77$, the first value extracted from
$r_{3,1}(d)$ and the second from $r_{2,2}(d)$, are consistent
with each other, although the statistical convergence of the two
two-point correlators of order $n_1+n_2=4$ is already beyond the
limit of acceptability. For the record w3 a similar statement can
be made, but the other records w1 and a are definitely confined to
$n_1+n_2 \leq 3$. Their best-fit parameters $L_\mathrm{casc}$,
$\tau_n$ and $c_n$ are also listed in Table II.

The cumulant $c_1$ cannot be directly extracted from two-point 
cumulants. However, an indirect extraction is possible via the 
relations (\ref{twoc7}) and (\ref{twoc8}). Using the numerical 
values of $c_2$ and $c_3$ determined already, the truncated 
multifractal sum rule (\ref{twoc7}) leads to the $c_1$ values 
listed in the second-last column of Table 2. Input for the
replica-trick formula (\ref{twoc8}) are the scaling exponents
$\tau_1$, $\tau_2$ and $\tau_3$ determined already. The
application of a cubic spline for the extrapolation results in the
last column of Table 2. If a linear spline is used instead, the
resulting $c_1$ values of the records w1, w2, w3 shift closer to
the values of the second-last column, but would do worse for the
record a. For an order-of-magnitude estimate, however, it is safe
to say that $c_1 \approx 0.05$ for the records w1, w2, w3 and $c_1
\approx 0.07$ for the record a.

This comparison of the RMCP theory with data demonstrates that the
parameter $L_{\rm casc}$ is a meaningful quantity and deserves
some fundamental consideration. For two-point distances
$20{-}30\eta \leq d \leq L_{\rm casc}$, the fitted expressions
(\ref{twob9}) and (\ref{twoc2}) are in good qualitative agreement
with their experimental counterparts. At $d=L_{\rm casc}$ all
two-point correlations decorrelate and become identical to their
asymptotic values corresponding to $d\to\infty$; $L_{\rm casc} >
L$ is somewhat larger than the operationally defined integral
length $L$ of Table I, which is calculated as the correlation
length of the velocity autocorrelation function. The physical
interpretation of the extracted parameter $L_{\rm casc}$ is that
of a turbulent cascade length which, according to Fig.\ 2,
describes the spatial extension of a hierarchical RMCP domain.

The finite-size scaling $f_{n_1,n_2}(d/L_{\rm casc})$ of two-point
correlators, predicted by RMCP, now allows for an unambiguous
derivation of scaling exponents, even for fully developed
turbulent flows with a moderate Reynolds number. Consequently, as
a closer inspection of Table II shows, reliable statements that the
intermittency exponent $\tau_2$ might show a weak dependence on
the Reynolds number appear to be within reach. Of course, to make
this statement solid, the analysis of many more records is needed;
this effort will be reported in \cite{CLE03a}.

%%%%%%%%%%%%%%%%%%%%%%%%%%%%%%%%%%%%%%%%%%%%%%%%%%%%%%%%%%%%%%%%%%%%%%%
%%%%%%%%%%%%%%%%%%%%%%%%%%%%%%%%%%%%%%%%%%%%%%%%%%%%%%%%%%%%%%%%%%%%%%%
\section{Parametric estimation of the RMCP generator}
%%%%%%%%%%%%%%%%%%%%%%%%%%%%%%%%%%%%%%%%%%%%%%%%%%%%%%%%%%%%%%%%%%%%%%%
%%%%%%%%%%%%%%%%%%%%%%%%%%%%%%%%%%%%%%%%%%%%%%%%%%%%%%%%%%%%%%%%%%%%%%%

If the scaling exponents $\tau_n$, or the cumulants $c_n$, exist and 
are known for all orders $1 \leq n < \infty$, the binary RMCP 
generator could in principle be reconstructed via the inverse 
transform of (\ref{twoc5}). Unfortunately, as we have seen in the 
previous section, reliable information is limited to the lowest 
orders. Hence, the best we can do is to use sophisticated parametric 
estimates. Section V.A lists some of the most popular parametrizations 
and compares their performance with the results listed in Table II. 
Section V.B introduces the so-called log-normal inverse Gaussian 
distribution \cite{BAR98,BAR01}, which represents a broader and more 
flexible parametrization class, with the purpose of finding a suitable 
approximation to the true cascade generator.

%%%%%%%%%%%%%%%%%%%%%%%%%%%%%%%%%%%%%%%%%%%%%%%%%%%%%%%%%%%%%%%%%%%%%%%
\subsection{Dictionary of prototype cascade generators}
%%%%%%%%%%%%%%%%%%%%%%%%%%%%%%%%%%%%%%%%%%%%%%%%%%%%%%%%%%%%%%%%%%%%%%%

Here we list a number of popular generators $p(q)$ for binary
random multiplicative cascade processes. They all have the
property that the expectation value $\langle q \rangle = 1$.

The log-normal distribution
%%%---------------------------------------------------------------
\begin{equation}
\label{fivea1}
  p_{\rm lognormal}(q)
    =  \frac{1}{\sqrt{2\pi} \sigma q}
       \exp\left[ - {1 \over 2 \sigma^2}
         \left( \ln q + \frac{\sigma^2}{2} \right)^2
       \right]
       \;
\end{equation}
%%%---------------------------------------------------------------
is classic \cite{OBU62,KOL62}. Its log-stable generalization has
also been considered \cite{KID91,SCH92}, but does not qualify for
our purposes, since the cumulants $c_n$ do not exist for this
distribution beyond some order. For comparison, we will also employ 
the rescaled gamma distribution
%%%---------------------------------------------------------------
\begin{equation}
\label{fivea2}
  p_{\rm gamma}(q)
    =  {\gamma^\gamma \over \Gamma(\gamma)}
       q^{\gamma-1} e^{-\gamma q}
\end{equation}
%%%---------------------------------------------------------------
and the asymmetric beta distribution \cite{JOU00}
%%%---------------------------------------------------------------
\begin{equation}
\label{fivea3}
  p_{\rm beta}(q)
    =  \frac{ \Gamma(8\beta) }
            { \Gamma(\beta)\Gamma(7\beta) } \;
       8^{1-8\beta}
       q^{\beta-1} (8-q)^{7\beta-1}
       \; .
\end{equation}
%%%---------------------------------------------------------------
The bimodal distribution
%%%---------------------------------------------------------------
\begin{equation}
\label{fivea4}
  p_{\rm alpha}(q)
    =  {\alpha_2 \over {\alpha_1+\alpha_2}}
       \delta\left( q-(1-\alpha_1) \right)
       + {\alpha_1 \over {\alpha_1+\alpha_2}}
       \delta\left( q-(1+\alpha_2) \right)
       \; ,
\end{equation}
%%%---------------------------------------------------------------
although discrete, has also been used extensively \cite{SCH85,MEN87}.
Another popular discrete representative is the log-Poisson
distribution
%%%---------------------------------------------------------------
\begin{equation}
\label{fivea5}
  p_{\rm logPoisson}(q)
    =  \sum_{k=0}^\infty
       \frac{ 2^{-\nu_1} (\nu_1 \ln{2})^k }{ k!}
       \delta\left(
       q - 2^{\nu_1(1-\nu_2)} \nu_2^k
       \right)
       \; ,
\end{equation}
%%%---------------------------------------------------------------
which was originally derived with $\nu_1=2$ and $\nu_2=2/3$ from
some plausible reasoning on the structure of the most singular
objects in fully developed turbulent flows
\cite{SHE94,DUB94,SHE95}.

For all parametrizations (\ref{fivea1})--(\ref{fivea5}) it is
straightforward to determine analytic expressions for the scaling
exponents and cumulants via (\ref{twoc4})--(\ref{twoc6}). The free 
parameter of the one-parametric distributions 
(\ref{fivea1})--(\ref{fivea3}) is then fixed to reproduce the 
observed intermittency exponent $\tau_2$, listed in Table II. The 
two-parametric distributions (\ref{fivea4}) and (\ref{fivea5}) 
need also conform to $\tau_3$ in addition to $\tau_2$. No further 
freedom is left for the scaling exponents of higher order and 
cumulants of all orders. Table III summarizes their predicted values.

It is difficult to rate the prototype cascade generators because
of ambiguity inherent in the data. Within the one-parametric
distributions (\ref{fivea1})-(\ref{fivea3}) the log-normal
distribution performs better: for all the records, the predicted
values for $c_1$ and $c_2$ are close to the observed cumulants.
However, the log-normal distribution without skewness is unable to 
reproduce the observed positive values for $c_3$ and for the record 
w2 it also overestimates the scaling exponent $\tau_4$. Furthermore, 
the difficulties of the log-normal distribution for high-order 
moments is now well known \cite{FRI95}. Compared to the log-normal 
distribution, the rescaled gamma distribution and the asymmetric 
beta distribution have the tendency to overestimate the first two 
cumulants. Furthermore, $c_3$ is predicted with an opposite sign. 
Rather surprisingly, the simple two-parametric bimodal distribution 
(\ref{fivea4}) shows the closest agreement for all records. The 
scaling exponent $\tau_4$, if observed, as well as the cumulants
$c_1$ and $c_2$ almost match their observed counterparts. Moreover, 
$c_3$ comes with the correct sign, although it is about a factor 
$2$ too low for the records w1, w2, w3 and roughly a factor $2$ too 
large for the atmospheric boundary layer record. Like the 
distributions (\ref{fivea2}) and (\ref{fivea3}), the two-parametric 
log-Poisson distribution overestimates the second cumulant $c_2$ 
and, except for record w3, predicts $c_3$ with the wrong sign. It 
is interesting to note that the parameter-free log-Poisson
distribution \cite{SHE94} with $\nu_1=2$ and $\nu_2=2/3$ matches
well the scaling exponents $\tau_3$ and $\tau_4$ of record a with
the largest Reynolds number, but disagrees with all cumulants.

%%%%%%%%%%%%%%%%%%%%%%%%%%%%%%%%%%%%%%%%%%%%%%%%%%%%%%%%%%%%%%%%%%%%%%%
\subsection{Log-normal inverse Gaussian distribution}
%%%%%%%%%%%%%%%%%%%%%%%%%%%%%%%%%%%%%%%%%%%%%%%%%%%%%%%%%%%%%%%%%%%%%%%

A broader and more flexible parametrization class is the so-called
normal inverse Gaussian distribution \cite{BAR98,BAR01}
%%%---------------------------------------------------------------
\begin{equation}
\label{fiveb1}
  p(x;\alpha ,\beta ,\mu ,\delta )
    =  a(\alpha ,\beta ,\mu ,\delta) \,
       s\!\left( \frac{x-\mu}{\delta } \right)^{-1}
       K_{1}\left\{
         \delta \alpha \, s\!\left( \frac{x-\mu}{\delta }\right)
       \right\}
       e^{\beta x}
       \; ,
\end{equation}
%%%---------------------------------------------------------------
with $s(x) = \sqrt{(1+x^{2})}$,
$a(\alpha,\beta,\mu,\delta) =
 \pi^{-1}\alpha \exp( \delta\sqrt{\alpha^2-\beta^2} - \beta\mu )$
and $-\infty{<}x{<}\infty$. $K_{1}$ is the modified Bessel
function of the third kind and index 1. The domain of variation of
the four parameters is given by $\mu \in \mathbf{R}$, $\delta \in
\mathbf{R}_{+}$ and $0\leq |\beta| <\alpha $. The distribution is
denoted by $\mathrm{NIG}(\alpha,\beta,\mu,\delta)$, and its
cumulant generating function
$\mathrm{Q}(\lambda;\alpha,\beta,\mu,\delta) =
 \ln\langle e^{\lambda x} \rangle$
has the simple form
%%%---------------------------------------------------------------
\begin{equation}
\label{fiveb2}
  \mathrm{Q}(\lambda;\alpha,\beta,\mu,\delta)
    =  \delta \left(
       \sqrt{\alpha^2-\beta^2}
       - \sqrt{\alpha^2-(\beta+\lambda)^2}
       \right)
       +\mu\lambda
       \; .
\end{equation}
%%%---------------------------------------------------------------
If $x_1,\ldots,x_m$ are independent normal inverse Gaussian random
variables with common parameters $\alpha$ and $\beta$ but
individual location-scale parameters $\mu_i$ and $\delta_i$
$(i=1,\ldots,m)$, then $x_{+}=x_1+\ldots+x_m$ is again distributed
according to a normal inverse Gaussian law with parameters
$\alpha$, $\beta$, $\mu_{+}$ and $\delta_{+}$. Furthermore, we
note that the $\mathrm{NIG}$ distribution (\ref{fiveb1}) has
semi-stretched tails
%%%---------------------------------------------------------------
\begin{equation}
\label{fiveb3}
  p(x;\alpha,\beta,\mu,\delta)
    \sim  \left\vert x \right\vert^{-3/2}
          \exp\left(
            -\alpha \left\vert x \right\vert + \beta x
          \right)
\end{equation}
%%%---------------------------------------------------------------
as $x\rightarrow\pm\infty$. This result follows from the
asymptotic relation $K_{\nu}(x{\rightarrow}\infty) \sim
\sqrt{\pi/2}x^{-1/2}e^{-x}$.

For our purposes we assume the random multiplicative weight to be
distributed according to
%%%---------------------------------------------------------------
\begin{equation}
\label{fiveb4}
  \ln q  \sim  \mathrm{NIG}(\alpha,\beta,\delta,\mu)
               \; ,
\end{equation}
%%%---------------------------------------------------------------
which turns normal inverse Gaussian statistics into log-normal
inverse Gaussian statistics. With (\ref{twoc4}), (\ref{twoc6}) and
(\ref{fiveb2}), the scaling exponents and cumulants yield
%%%---------------------------------------------------------------
\begin{equation}
\label{fiveb5}
  \tau_n \ln{2}
    =  \mathrm{Q}(n;\alpha,\beta,\mu,\delta)
\end{equation}
%%%---------------------------------------------------------------
and
%%%---------------------------------------------------------------
\begin{equation}
\label{fiveb6}
  c_1  =  \mu +\frac{\delta\rho}{\sqrt{1-\rho^2}}
          \; , \quad
  c_2  =  \frac{\delta}{\alpha(1-\rho^2)^{3/2}}
          \; , \quad
  c_3  =  \frac{3\delta\rho}{\alpha^2(1-\rho^2)^{5/2}}
          \; , \quad \ldots \; ,
\end{equation}
%%%---------------------------------------------------------------
where $\rho=\beta/\alpha$.

For each of the records w1, w2, w3 and a, the four NIG parameters
$\alpha$, $\beta$, $\delta$ and $\mu$ are determined so as to
reproduce $\tau_1=0$ and the observed values for $\tau_2$,
$\tau_3$ and $c_2$ listed in Table \ref{table2}. Since the
respective expressions (\ref{fiveb5}) and (\ref{fiveb6}) are
nonlinear, real solutions for the parameters are not guaranteed.
Where complex-valued solutions resulted in the first attempt,
which occurred for w2, w3 and a, the values for $c_2$, $\tau_3$
and $\tau_2$ are relaxed, in this order and to some small extent,
until real-valued parameter solutions are obtained. The outcome is
listed in Table \ref{table4}. The results are very close to the
log-normal values listed in \ref{table3}. The log-normal inverse
Gaussian distribution has the tendency to overestimate the
fourth-order scaling exponent $\tau_4$. The magnitude of the third
cumulant $c_3$ is strongly underestimated, so that its predicted 
sign shows only random scatter. Figure \ref{figure6} compares the 
extracted log-normal inverse Gaussian distributions of Table 
\ref{table4} with the extracted log-normal distributions of Table 
\ref{table3}. For all four records the extracted distributions are 
very similar.

As a summary of this section, we reiterate that the bimodal
distribution (\ref{fivea4}) produces the best overall agreement
with the observed scaling exponents and cumulants. However, the true 
cascade generator will not be discrete. From the set of continuous 
generator representatives tested, the log-normal and log-normal 
inverse Gaussian distributions perform best and about equally well.

%%%%%%%%%%%%%%%%%%%%%%%%%%%%%%%%%%%%%%%%%%%%%%%%%%%%%%%%%%%%%%%%%%%%%%%
%%%%%%%%%%%%%%%%%%%%%%%%%%%%%%%%%%%%%%%%%%%%%%%%%%%%%%%%%%%%%%%%%%%%%%%
\section{Conclusion}
%%%%%%%%%%%%%%%%%%%%%%%%%%%%%%%%%%%%%%%%%%%%%%%%%%%%%%%%%%%%%%%%%%%%%%%
%%%%%%%%%%%%%%%%%%%%%%%%%%%%%%%%%%%%%%%%%%%%%%%%%%%%%%%%%%%%%%%%%%%%%%%

Random multiplicative cascade processes are able to describe the
observed two-point correlation structure of the surrogate energy
dissipation of fully developed turbulent flows beyond simple
power-law scaling. Keeping in mind the need for a satisfactory
comparison between modelling and experimental data, a useful
transformation has been introduced: this transformation converts 
model-dependent and unobservable ultrametric two-point
distances to Euclidean two-point distances, the latter reflecting
the horizontally sampled $n$-point statistics of the experimental
records. The predictions of RMCP for finite-size scaling of
two-point correlation functions are confirmed by experimental data
from three (wind tunnel) shear flows and one atmospheric boundary
layer; a new physical length scale characterizing the upper end of
the inertial range, called here the turbulent cascade length, has
been shown to appear naturally. Furthermore, the quantitative
classification of the deviations from a rigorous scaling of
two-point correlators allows for an unambiguous extraction of
multiscaling exponents, even for flows with moderate Reynolds
numbers. When complemented with additional information extracted
from two-point cumulants of the logarithmic energy dissipation,
reliable parametric estimates of the RMCP generator have become
feasible.

RMCPs produce a consistent geometrical modelling of the self-similar
turbulent energy cascade. This is further supported by recent
investigations on the scaling part of three-point statistics
\cite{SCH03} and previous investigations on scale correlations
\cite{JOU99,JOU00,JOU02,CLE00}. Thus, the self-similar and
RMCP-like cascade process appears to be universal. However, as
shown by variations among the four records considered, the
strength of the cascade generator appears to depend on the
Reynolds number and perhaps also the flow geometry. This needs
further clarification by means of a separate and extended effort
\cite{CLE03a}; needless to say, there is a strong need for the
experimentalists to produce clean and longer records with
converged statistics.

\acknowledgements
The authors acknowledge fruitful discussions with Hans C.\ Eggers
and Markus Abel.

\newpage
%%%%%%%%%%%%%%%%%%%%%%%%%%%%%%%%%%%%%%%%%%%%%%%%%%%%%%%%%%%%%%%%%%%%%%%%%=

\newpage
%%%%%%%%%%%
% TABLE 1 %
%%%%%%%%%%%
\begin{table}
 \begin{tabular}{l|ccccc}
   \hline
   data set & $R_{\lambda}$ & $L/\eta$ & $L_{\rm record}/L$
   & $\lambda/\eta$ & $\Delta x/\eta$ \\
   \hline
   w1  & $306$   & $484$             & $102500$ & $35$  & $1.97$\\
   w2  & $493$   & $968$             & $193500$ & $44$  & $2.79$\\
   w3  & $1045$  & $2564$            & $77500$  & $64$  & $2.97$\\
   a   & $9000$  & $5{\times}10^4$   & $1000$   & $187$ & $1.29$\\
   \hline
 \end{tabular}
 \caption{
  Taylor-scale based Reynolds number $R_{\lambda}$,
  integral length scale $L$ in units of the dissipation scale $\eta$,
  record length $L_{\rm record}$,
  Taylor microscale $\lambda$ and
  resolution scale $\Delta x$ of
  three wind tunnel (w1, w2, w3) \cite{PEA02} and
  one atmospheric boundary layer (a) \cite{DHR00} records.
 }
 \label{table1}
\end{table}

%%%%%%%%%%%
% TABLE 2 %
%%%%%%%%%%%
\begin{table}
\begin{tabular}{c|c|ccccc|cc}
   \hline
   data set & $L_{\rm casc}/\eta$ & $\tau_2$ & $\tau_3$ & $\tau_4$
     & $c_2$ & $c_3$ & $c_1$(\ref{twoc7}) & $c_1$(\ref{twoc8}) \\
   \hline
   w1 & 1873   & 0.15 & 0.46 & --   & 0.100 & 0.042 & -0.057 &  -0.045 \\
   w2 & 3069   & 0.15 & 0.42 & 0.78 & 0.095 & 0.045 & -0.055 &  -0.046 \\
   w3 & 7117   & 0.17 & 0.52 & 0.98 & 0.099 & 0.060 & -0.059 &  -0.046 \\
   a  & 322500 & 0.21 & 0.58 & --   & 0.149 & 0.015 & -0.077 &  -0.068 \\
   \hline
\end{tabular}
\caption{ Parameter values resulting from least-square fits with
expressions (\ref{twob9}) and (\ref{twoc2}). $L_{\rm casc}$ has
been fixed for each data set. The last two columns represent the
estimates (\ref{twoc7}) and (\ref{twoc8}) for the cumulant $c_1$. } 
\label{table2}
\end{table}

%%%%%%%%%%%
% TABLE 3 %
%%%%%%%%%%%
\begin{table}
\begin{tabular}{l|c|ccccc}
  \hline
  \multicolumn{7}{c}{log-normal distribution (\ref{fivea1})}\\ \hline
  data set & $\sigma$ &
  $\tau_3$ & $\tau_4$ & $c_1$ & $c_2$ & $c_3$ \\
  \hline
  w1 & 0.33 & 0.46 & 0.93 & -0.054 & 0.107 & 0.000 \\
  w2 & 0.32 & 0.44 & 0.87 & -0.050 & 0.101 & 0.000 \\
  w3 & 0.34 & 0.51 & 1.03 & -0.059 & 0.119 & 0.000 \\
  a  & 0.38 & 0.63 & 1.26 & -0.073 & 0.145 & 0.000 \\
  \hline
  \multicolumn{7}{c}{gamma distribution (\ref{fivea2})}\\ \hline
  data set & $\gamma$ &
  $\tau_3$ & $\tau_4$ & $c_1$ & $c_2$ & $c_3$ \\
  \hline
  w1 & 8.84 & 0.45 & 0.87 & -0.058 & 0.120 & -0.014 \\
  w2 & 9.41 & 0.42 & 0.82 & -0.054 & 0.112 & -0.013 \\
  w3 & 7.94 & 0.50 & 0.96 & -0.064 & 0.134 & -0.018 \\
  a  & 6.39 & 0.60 & 1.16 & -0.080 & 0.169 & -0.029 \\
  \hline
  \multicolumn{7}{c}{beta distribution (\ref{fivea3})}\\  \hline
  data set & $\beta$ &
  $\tau_3$ & $\tau_4$ & $c_1$ & $c_2$ & $c_3$ \\
  \hline
  w1 & 7.61 & 0.44 & 0.85 & -0.059 &  0.124 & -0.019 \\
  w2 & 8.11 & 0.42 & 0.81 & -0.055 &  0.116 & -0.017 \\
  w3 & 6.82 & 0.49 & 0.94 & -0.066 &  0.139 & -0.025 \\
  a  & 5.47 & 0.60 & 1.13 & -0.083 &  0.178 & -0.040 \\
  \hline
  \multicolumn{7}{c}{bimodal distribution (\ref{fivea4})}\\ \hline
  data set & \multicolumn{1}{c}{$\alpha_1$} =
&\multicolumn{1}{c|}{$\alpha_2$} &
  $\tau_4$ & $c_1$ & $c_2$ & $c_3$ \\
  \hline
  w1 & \multicolumn{1}{c}{0.22} & \multicolumn{1}{c|}{0.52}
     & 0.88 & -0.049 & 0.092 & 0.025 \\
  w2 & \multicolumn{1}{c}{0.24} & \multicolumn{1}{c|}{0.44}
     & 0.80 & -0.049 & 0.094 & 0.018 \\
  w3 & \multicolumn{1}{c}{0.21} & \multicolumn{1}{c|}{0.61}
     & 1.00 & -0.052 & 0.094 & 0.033 \\
  a  & \multicolumn{1}{c}{0.31} & \multicolumn{1}{c|}{0.50}
     & 1.06 & -0.075 & 0.144 & 0.026 \\
  \hline
  \multicolumn{7}{c}{log-Poisson distribution (\ref{fivea5})
                     with $\nu_1=2$, $\nu_2=2/3$ }\\ \hline
  & \multicolumn{1}{c}{$\tau_2$}& $\tau_3$ & $\tau_4$ & $c_1$ & $c_2$ & $c_3$\\
  \hline
  & \multicolumn{1}{c}{0.22} & 0.59 & 1.06 & -0.10 & 0.228 & -0.092 \\
  \hline
  \multicolumn{7}{c}{log-Poisson distribution (\ref{fivea5})}\\ \hline
  data set & \multicolumn{1}{c}{$\nu_1$} & \multicolumn{1}{c|}{$\nu_2$}
     & $\tau_4$ & $c_1$ & $c_2$ & $c_3$ \\
  \hline
  w1 & \multicolumn{1}{c}{109.84} & \multicolumn{1}{c|}{0.96}
     & 0.90 & -0.055 & 0.111 & -0.004 \\
  w2 & \multicolumn{1}{c}{ 13.75} & \multicolumn{1}{c|}{0.90}
     & 0.82 & -0.054 & 0.112 & -0.012 \\
  w3 & \multicolumn{1}{c}{1548.7} & \multicolumn{1}{c|}{1.01}
     & 1.03 & -0.059 & 0.117 &  0.001 \\
  a  & \multicolumn{1}{c}{  4.61} & \multicolumn{1}{c|}{0.79}
     & 1.09 & -0.085 & 0.184 & -0.044 \\
  \hline
\end{tabular}
\caption{ Fitted parameters for a few prototype cascade generators
and their predicted values for the remaining scaling exponents
$\tau_n$ and cumulants $c_n$. }
\label{table3}
\end{table}

%%%%%%%%%%%
% TABLE 4 %
%%%%%%%%%%%
\begin{table}
\begin{tabular}{l|cccc|cccccc}
  \hline
  \multicolumn{11}{c}{log-normal inverse Gaussian distribution
                      (\ref{fiveb4})}\\
  \hline
  data set & $\alpha$ & $\beta$  & $\delta$ & $\mu$
           & $\tau_2$ & $\tau_3$ & $\tau_4$ & $c_1$  & $c_2$ & $c_3$  \\
  w1       & 10.98    & 0.94     & 1.09     & -0.14
           & 0.15     & 0.46     & 0.95     & -0.051 & 0.100 & 0.002  \\
  w2       & 17.28    & -5.46    & 1.56     & 0.47
           & 0.15     & 0.43     & 0.85     & -0.052 & 0.106 & -0.006 \\
  w3       & 27.26    & 17.80    & 1.17     & -1.06
           & 0.16     & 0.52     & 1.11     & -0.052 & 0.099 & 0.012  \\
  a        & 6.99     & -2.44    & 0.92     & 0.27
           & 0.20     & 0.60     & 1.19     & -0.076 & 0.160 & -0.027 \\
  \hline
\end{tabular}
\caption{
Fitted parameters of the log-normal inverse Gaussian cascade
generator and their predicted values for the scaling exponents
$\tau_n$ and cumulants $c_n$.
}
\label{table4}
\end{table}

\newpage
%%%%%%%%%%%%%
% FIGURE 1  %
%%%%%%%%%%%%%
\begin{figure}
\begin{centering}
\includegraphics[width=0.49\linewidth]{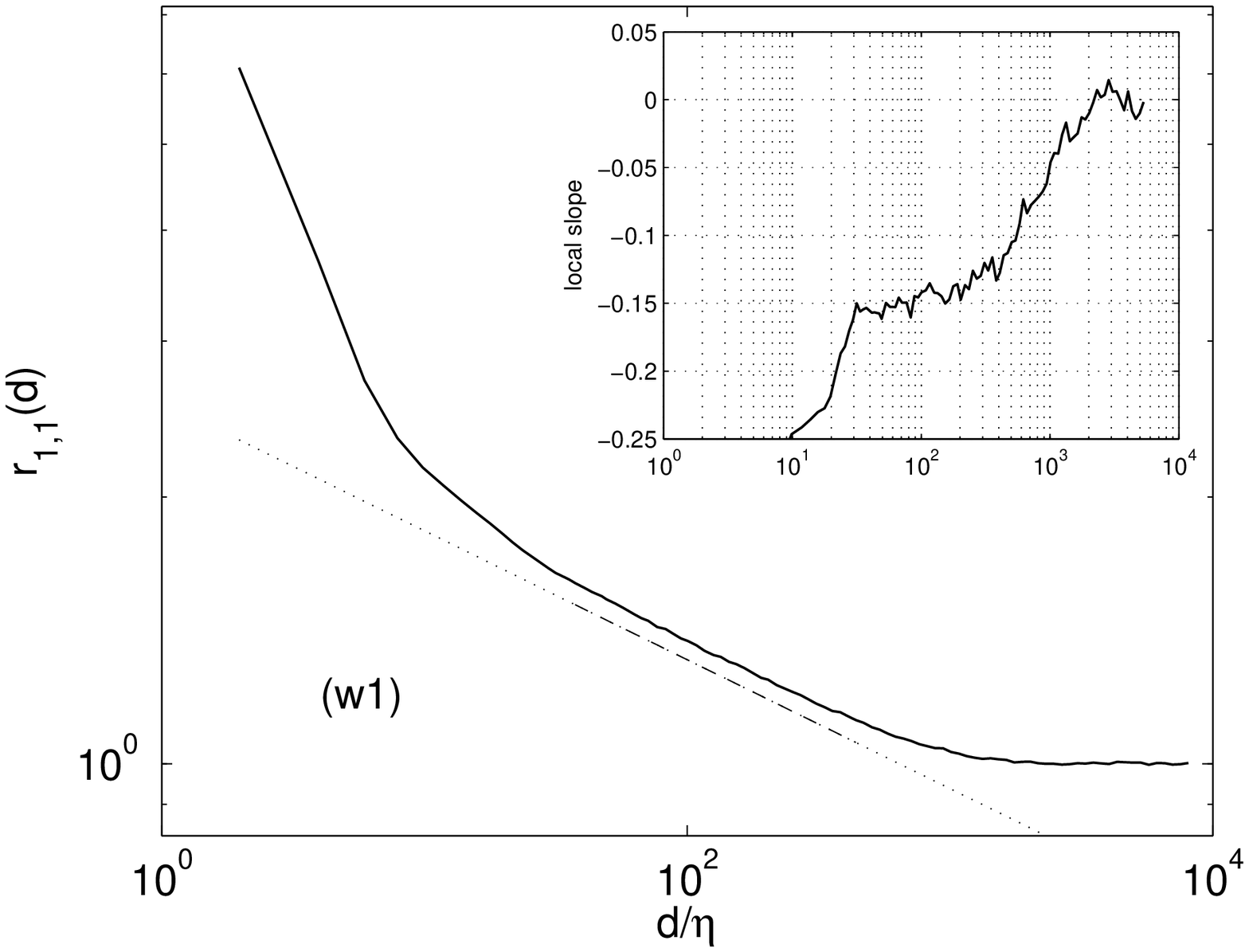}
\includegraphics[width=0.49\linewidth]{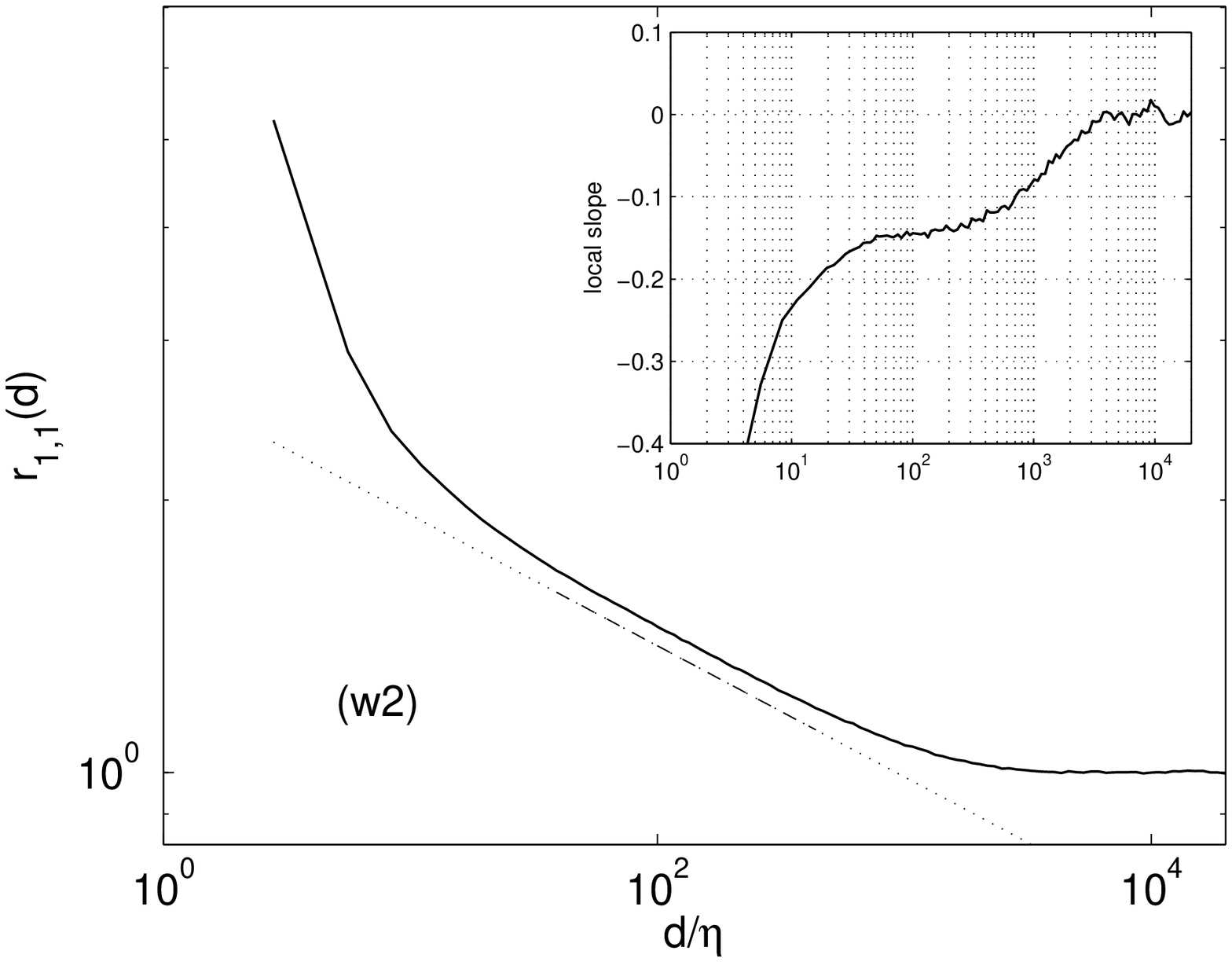}
\includegraphics[width=0.49\linewidth]{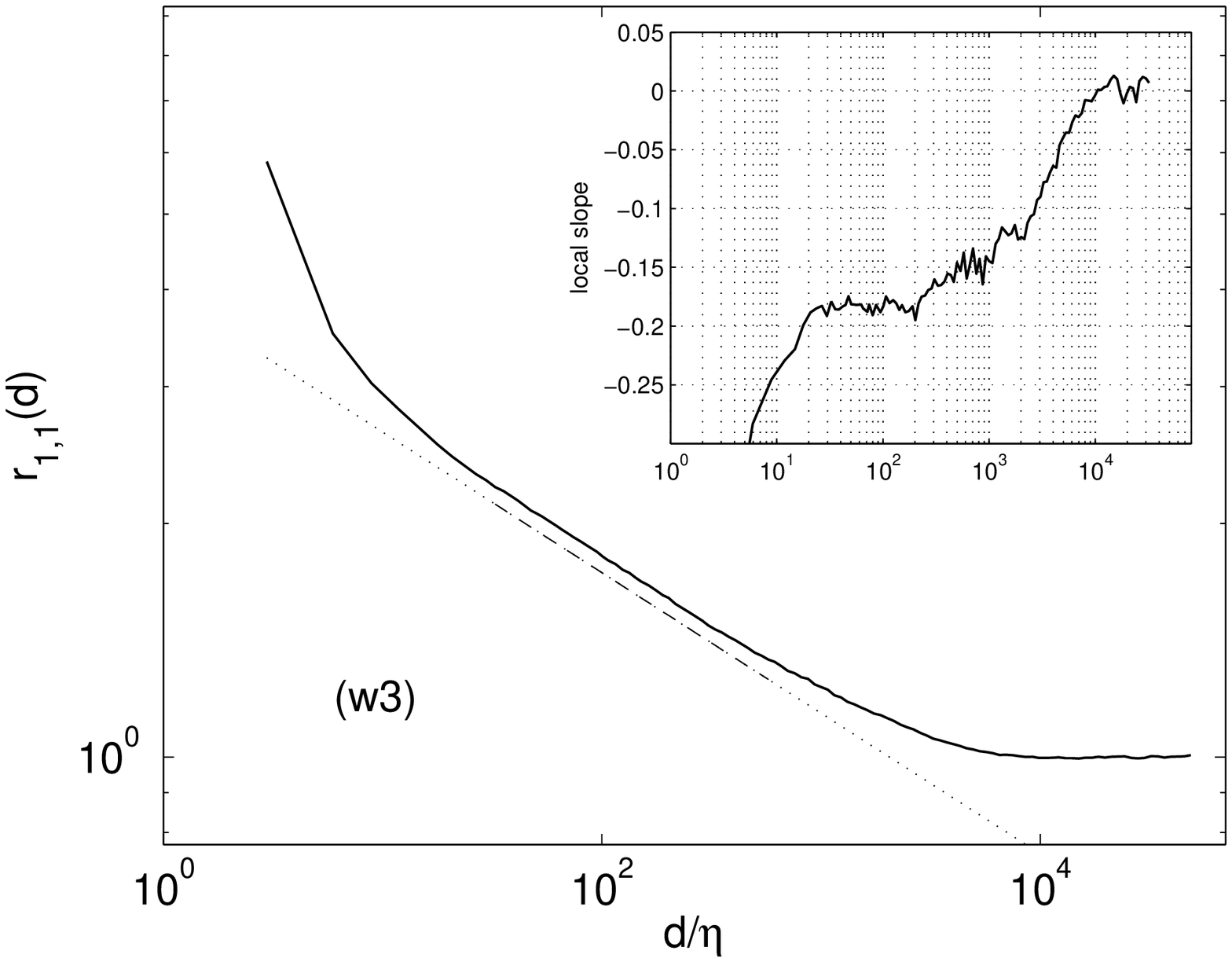}
\includegraphics[width=0.49\linewidth]{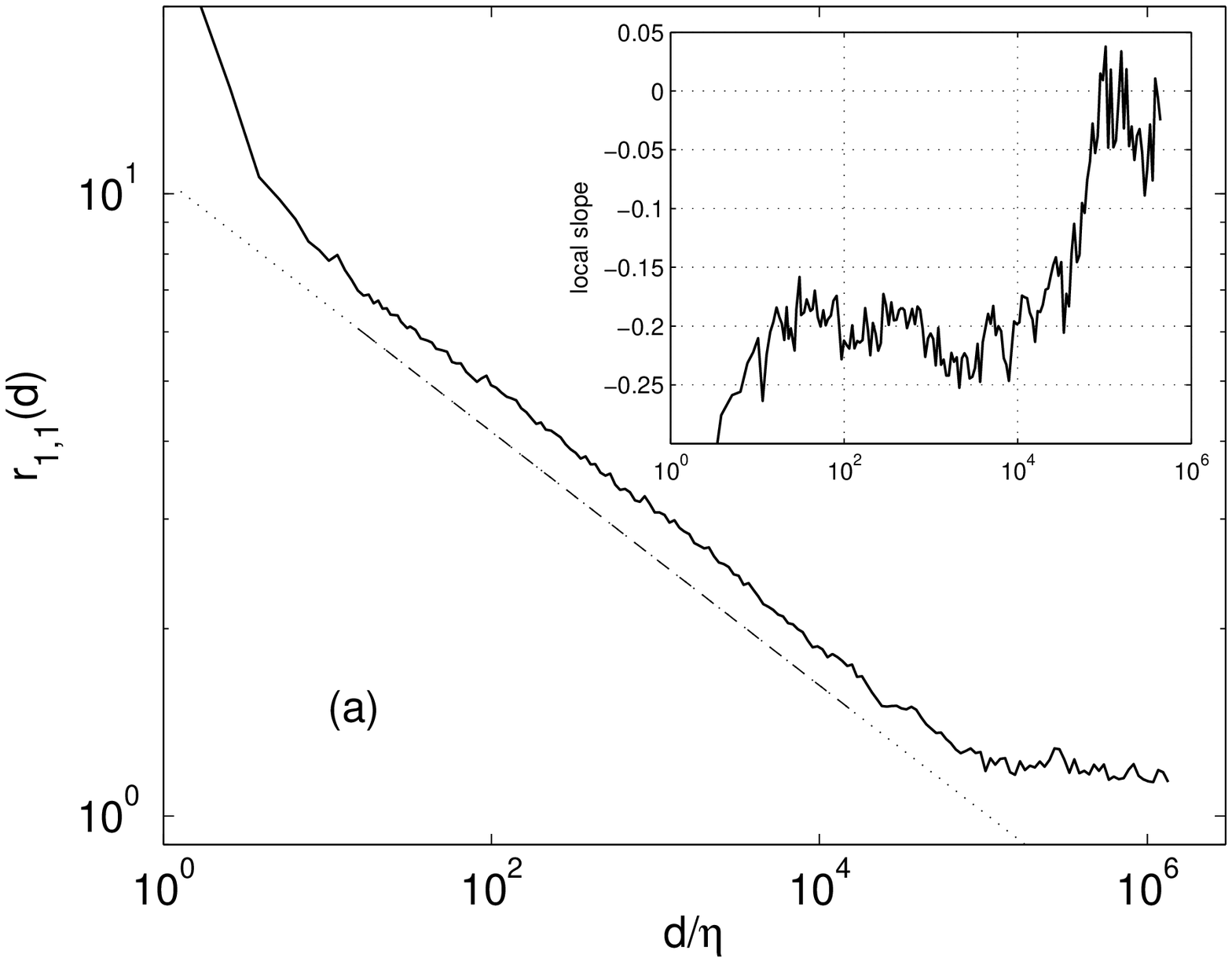}
\caption{
Two-point correlator
$\langle \varepsilon(x+d) \varepsilon(x) \rangle /
\langle \varepsilon(x) \rangle^2$
of the surrogate energy dissipation extracted from the records
w1, w2, w3 and a. Power-law fits are indicated by the shifted
broken straight lines. The inset figures illustrate the local
slope.
}
\label{figure1}
\end{centering}
\end{figure}

\newpage
%%%%%%%%%%%%%
% FIGURE 2  %
%%%%%%%%%%%%%
\begin{figure}
\begin{centering}
\includegraphics[width=\linewidth]{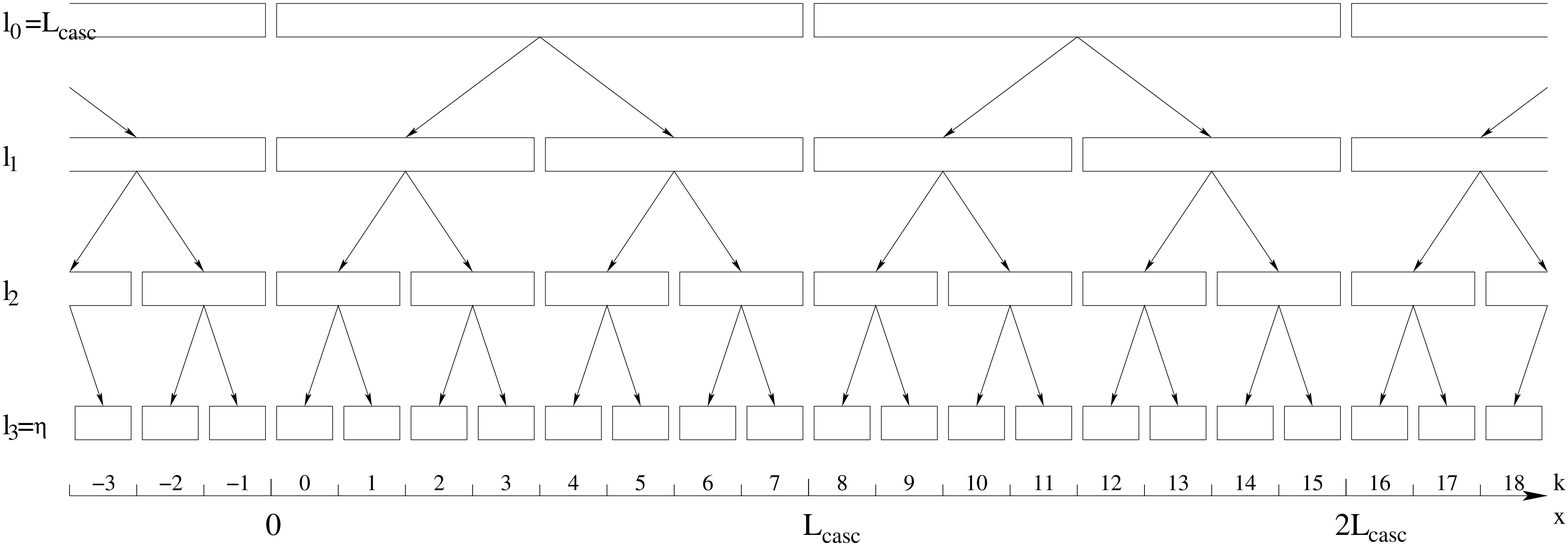}
\caption{
A chain of hierarchical RMCP domains of equal cascade length
$L_{\rm casc}$ is employed to convert the ultrametric two-point
statistics into an Euclidean one.
}
\label{figure2}
\end{centering}
\end{figure}

\newpage
%%%%%%%%%%%%%
% FIGURE 3  %
%%%%%%%%%%%%%
\begin{figure}
\begin{centering}
\includegraphics[width=\linewidth]{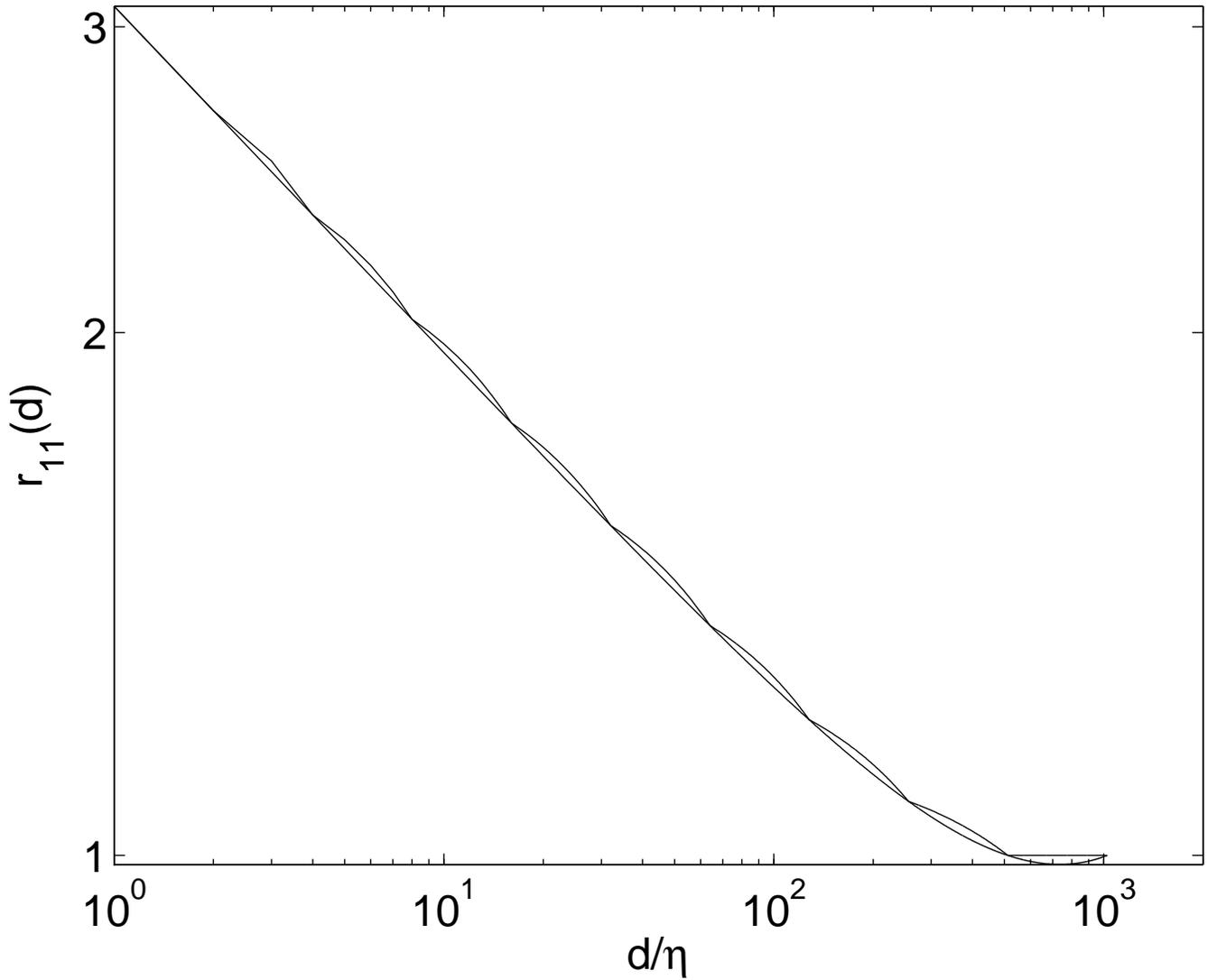}
\caption{ Comparison of the expressions (\ref{twob8}) and
(\ref{twob9}) for the order $n_1=n_2=1$, showing that the
log-oscillations inherent to (\ref{twob8}) remain small.
Parameters have been set as follows: $L_{\rm casc}/\eta=2^{10}$,
$\tau_2=0.20$ and $\Pi=1$. } 
\label{figure3}
\end{centering}
\end{figure}

\newpage
%%%%%%%%%%%%%
% FIGURE 4  %
%%%%%%%%%%%%%
\begin{figure}
\centering
\includegraphics[width=0.49\linewidth]{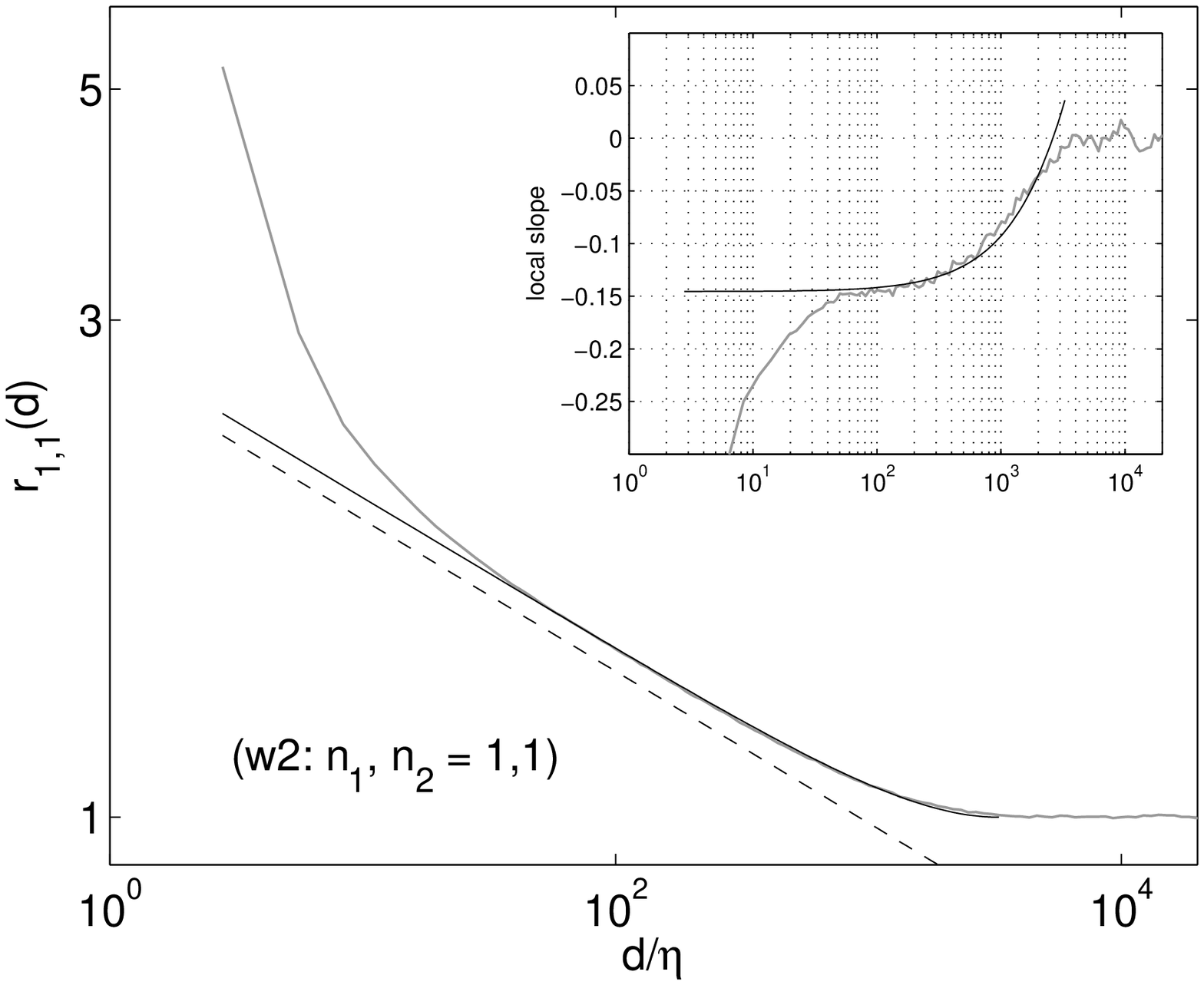}
\includegraphics[width=0.49\linewidth]{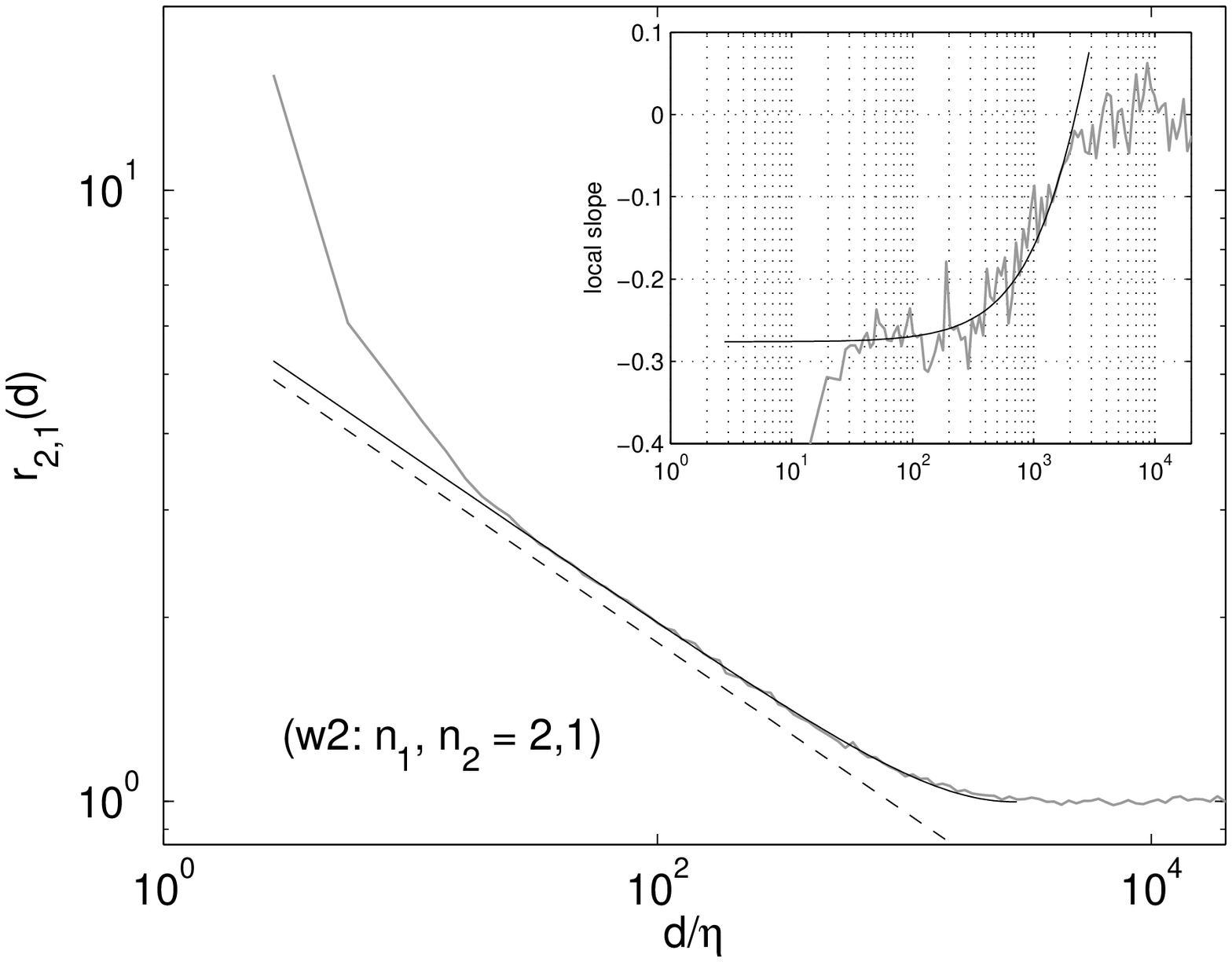}
\includegraphics[width=0.49\linewidth]{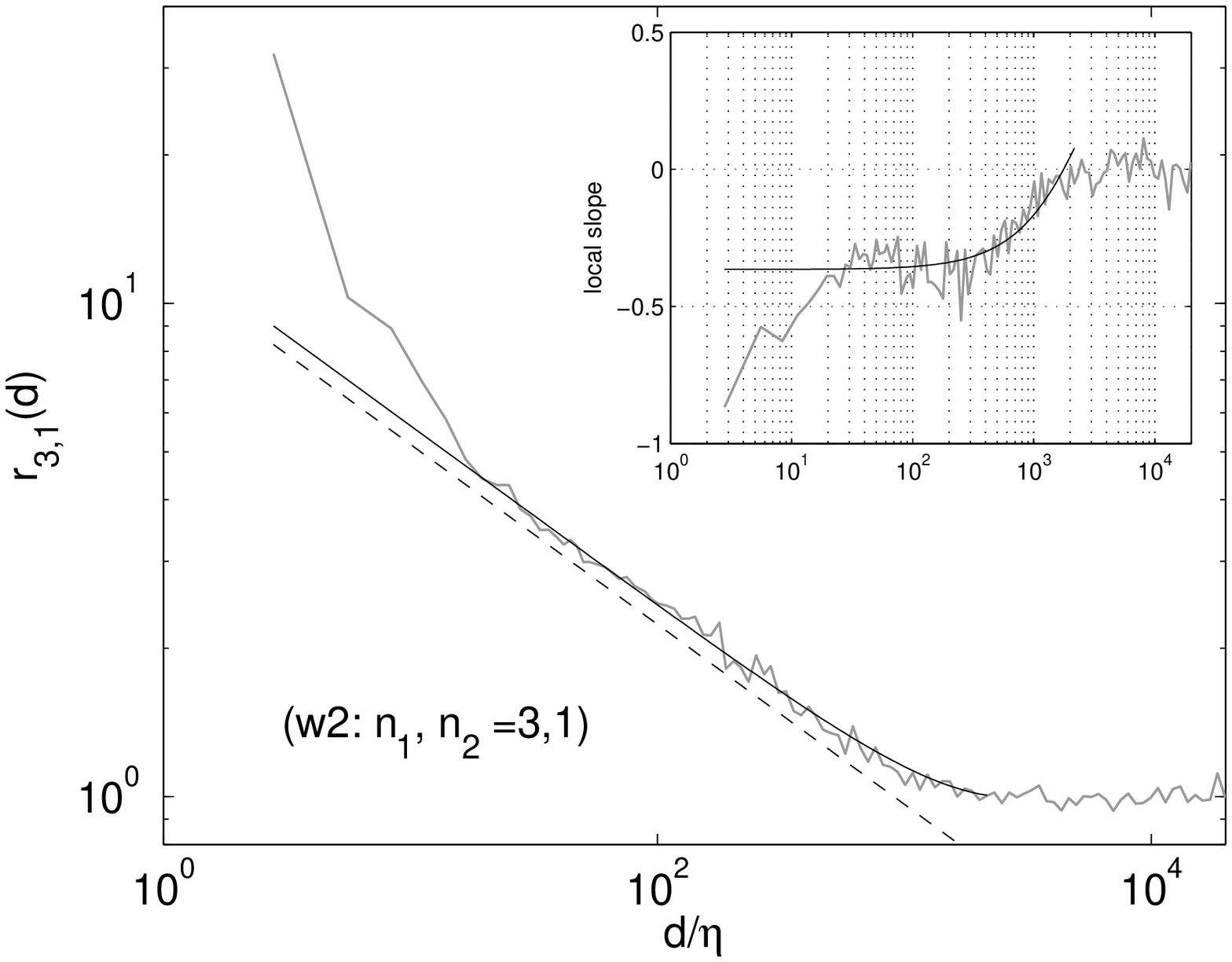}
\includegraphics[width=0.49\linewidth]{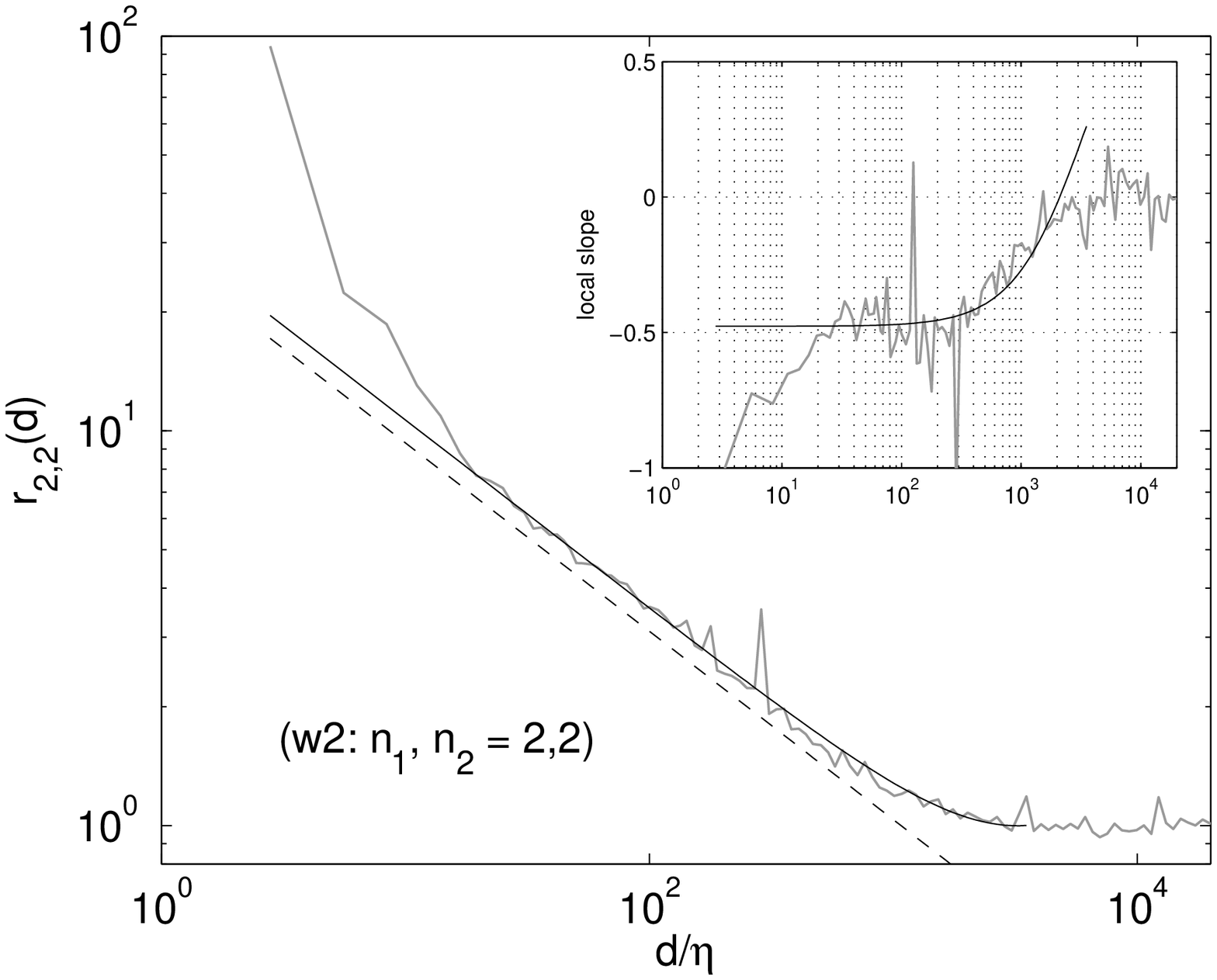}
\caption{ Best fits of expression (\ref{twob9}) to two-point
correlators extracted from data set w2. Correlation orders are
$n_1,n_2=1,1$, $2,1$, $3,1$ and $2,2$. The parameter $L_{\rm
casc}$ has been fitted such that it is the same for all orders of
two-point correlators as well as two-point cumulants, with the
latter illustrated in Fig.\ 5. The inset figures illustrate the
local slope. For comparison, power-law fits with the extracted
scaling exponents listed in Tab.\ II are shown as the shifted
dashed straight lines. } \label{figure4}
\end{figure}

\newpage
%%%%%%%%%%%%%
% FIGURE 5  %
%%%%%%%%%%%%%
\begin{figure}
\centering
\includegraphics[width=0.49\linewidth]{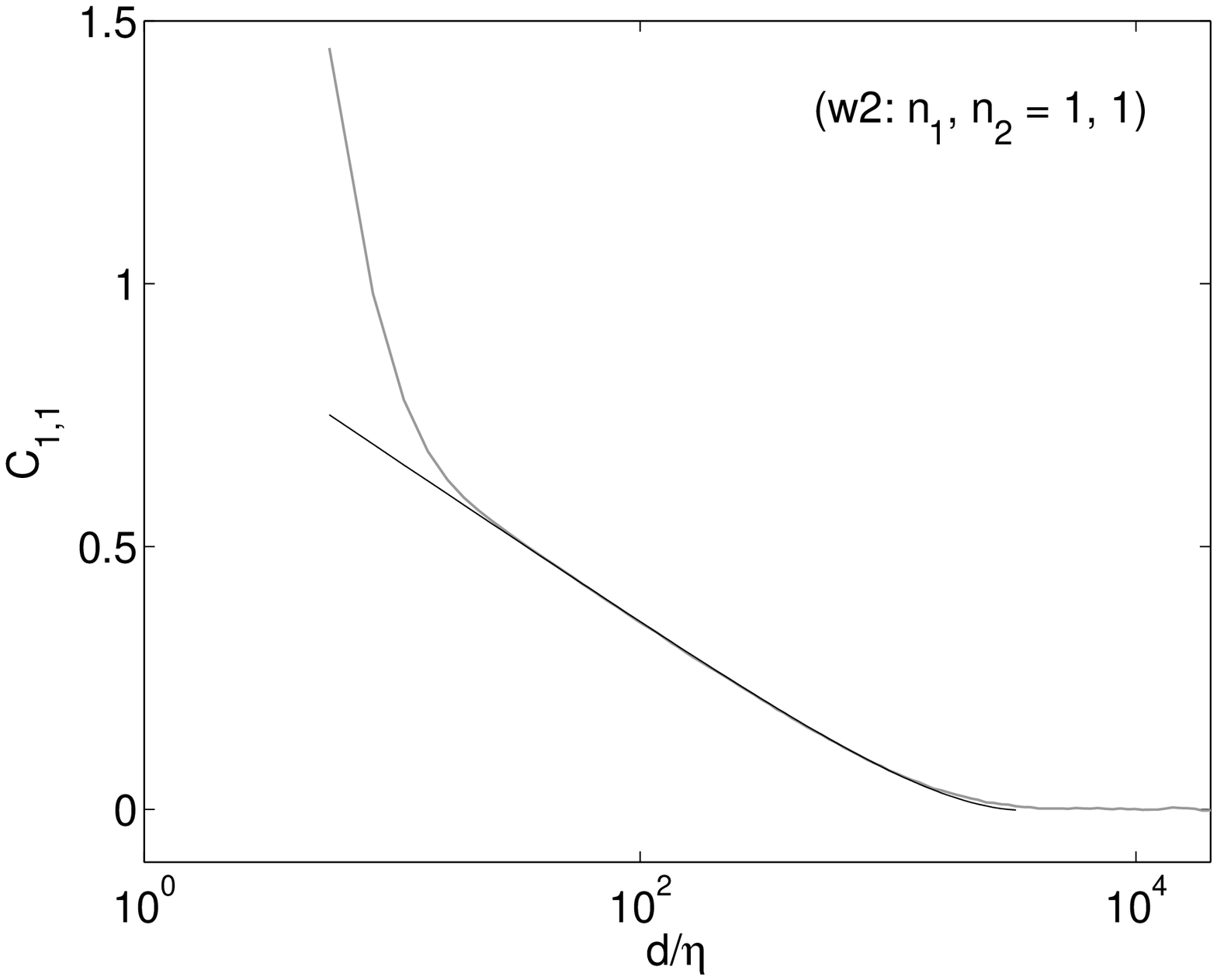}
\includegraphics[width=0.49\linewidth]{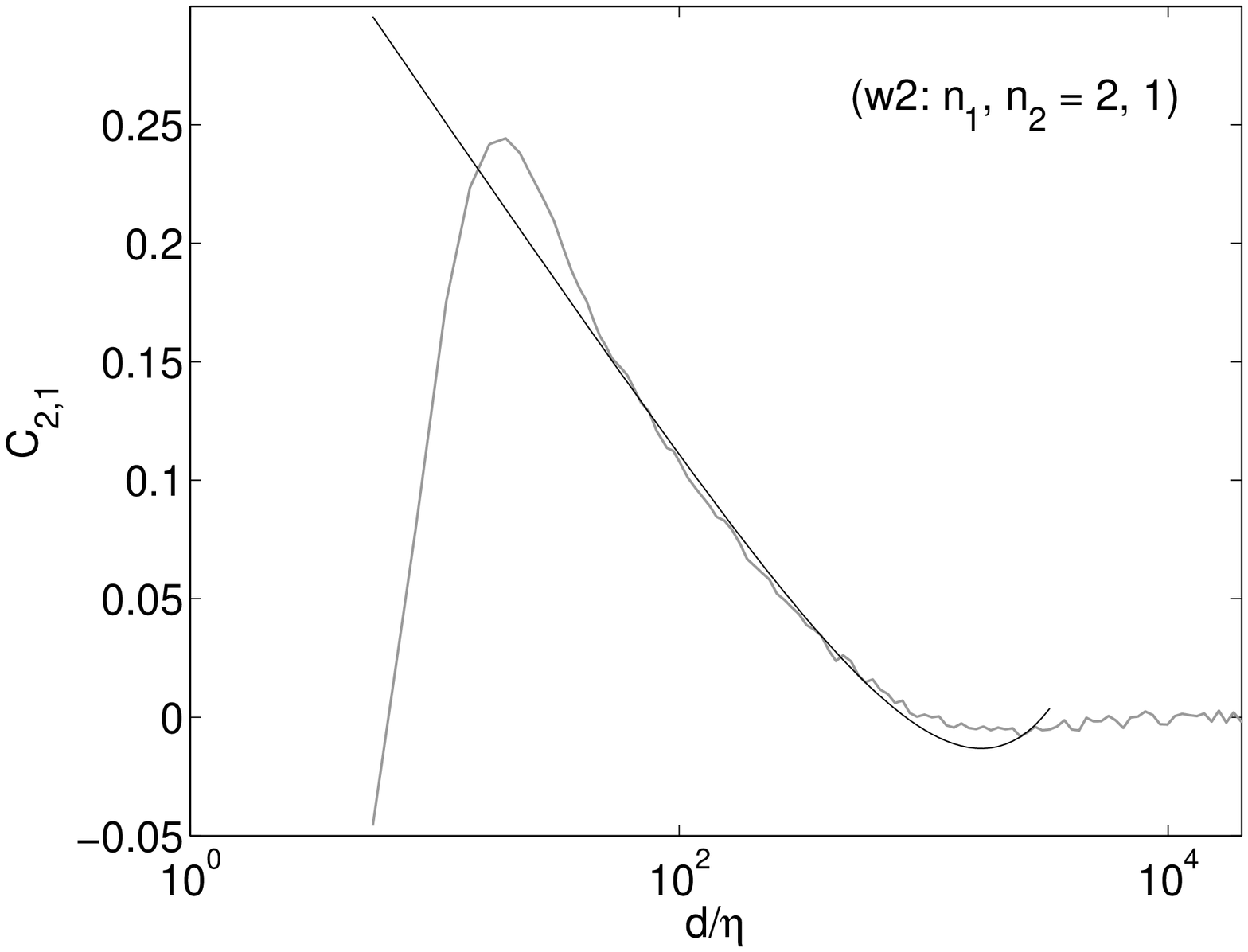}
\caption{ Best fits of expressions (\ref{twoc2}) to two-point
cumulants extracted from data set w2.  Correlation orders are
$n_1,n_2=1,1$ and $2,1$. The parameter $L_{\rm casc}$ has been
fitted such that it is the same for all the orders of two-point
cumulants as well as two-point correlators, with the latter
illustrated in Fig.\ 4. } \label{figure5}
\end{figure}

\newpage
%%%%%%%%%%%%%
% FIGURE 6  %
%%%%%%%%%%%%%
\begin{figure}
\centering
\includegraphics[width=0.49\linewidth]{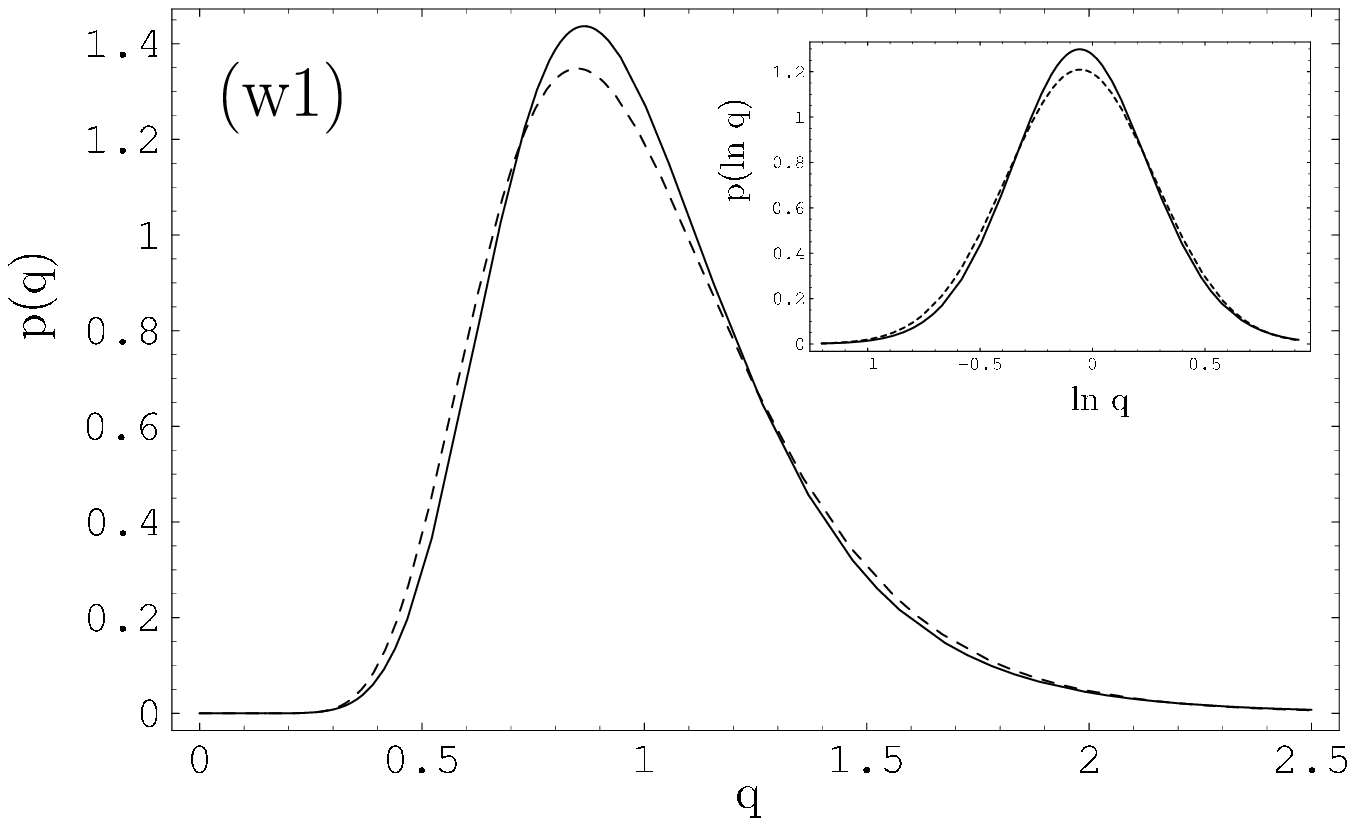}
\includegraphics[width=0.49\linewidth]{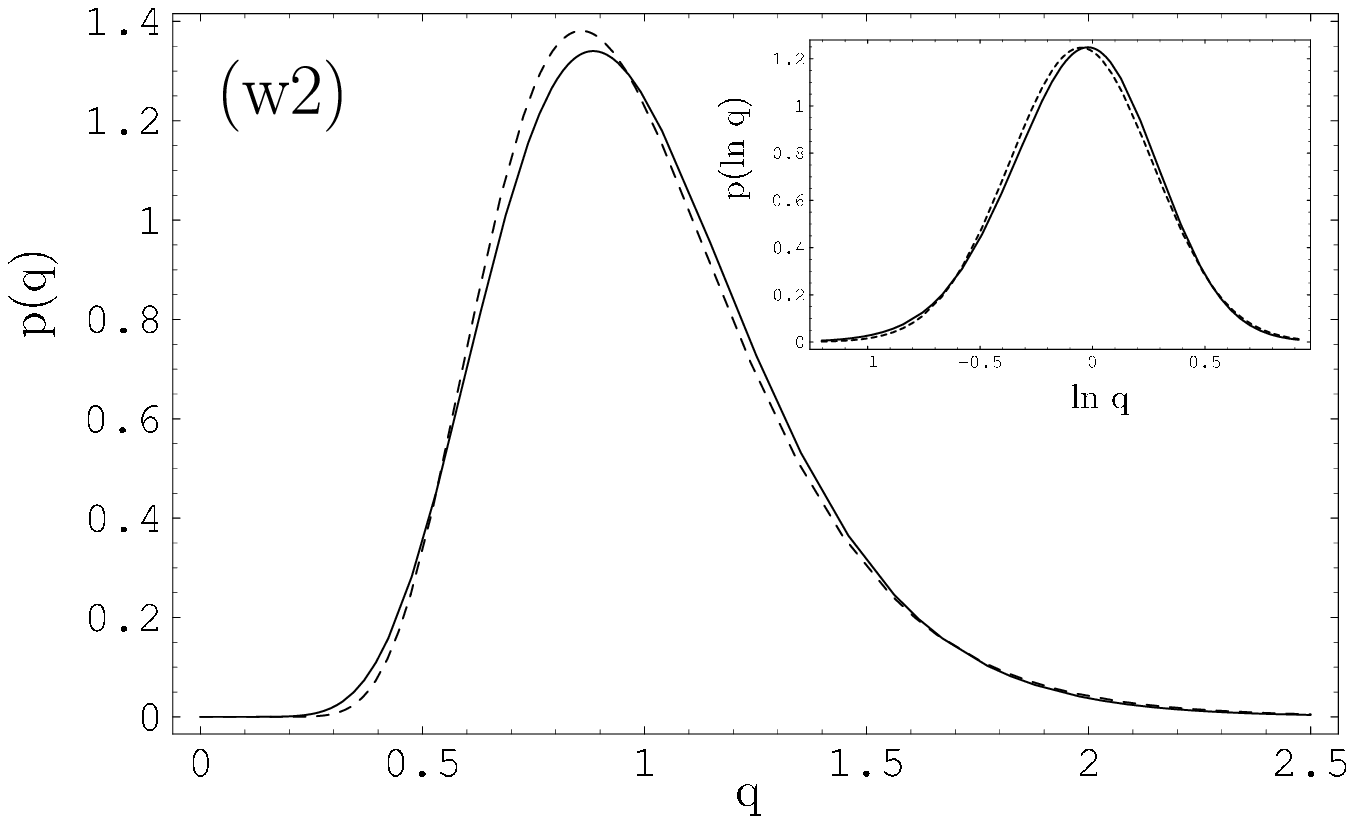}
\includegraphics[width=0.49\linewidth]{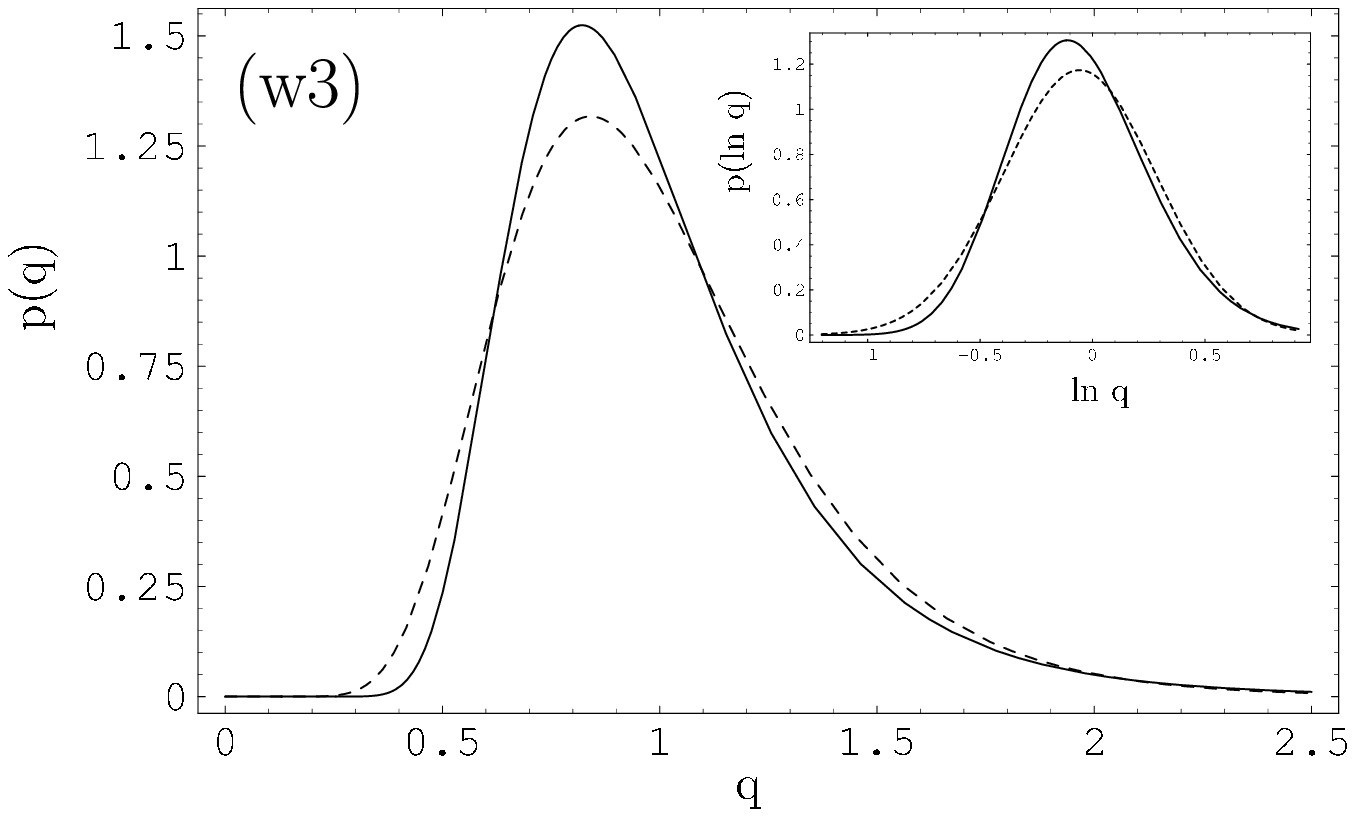}
\includegraphics[width=0.49\linewidth]{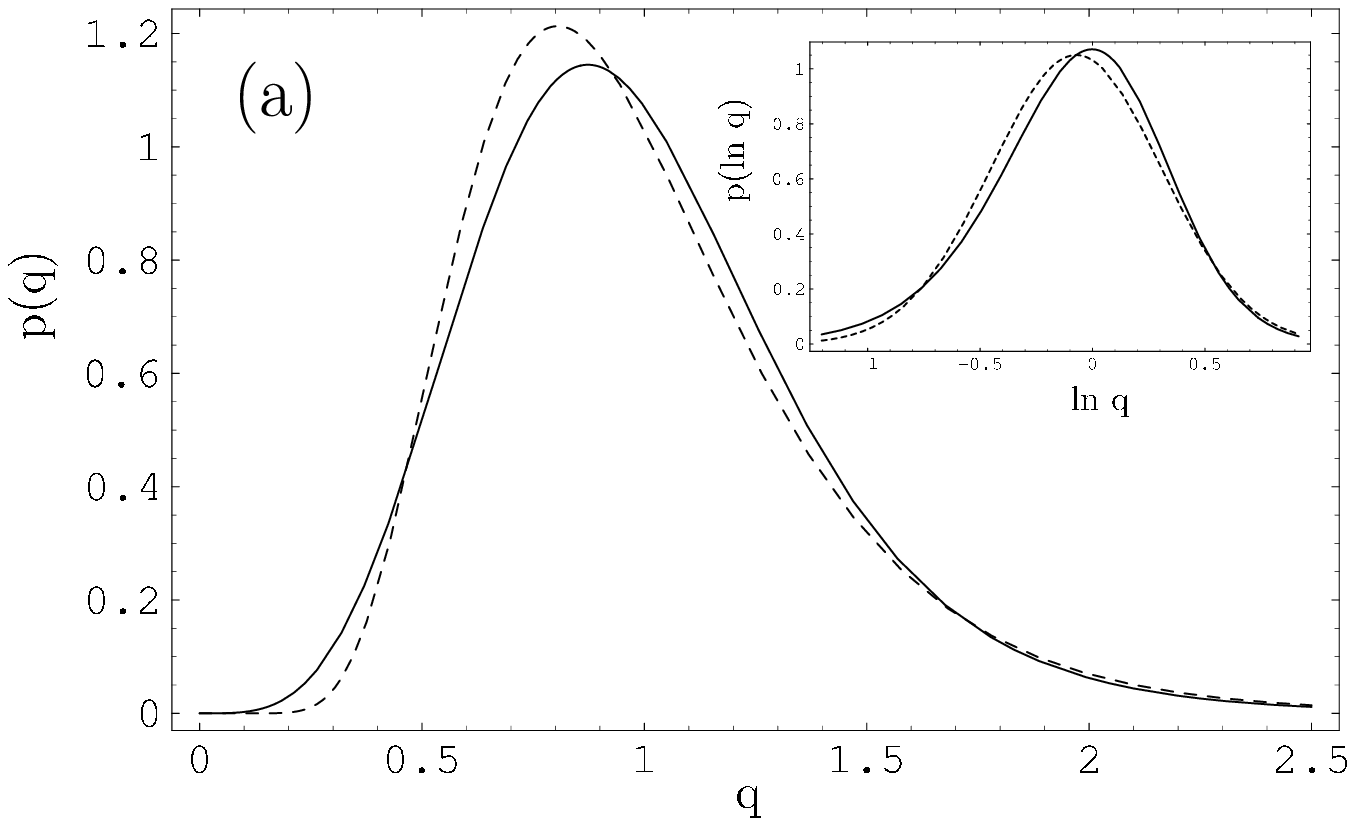}
\caption{
Comparison of the extracted log-normal distribution (\ref{fivea1})
(solid) and log-normal inverse Gaussian distribution (\ref{fiveb4}) 
(dashed) for the records w1, w2, w3 and a. Respective parameter 
values have been taken from Tables \ref{table3} and \ref{table4}.
}
\label{figure6}
\end{figure}

\end{document}